\documentclass[twocolumn,aps,showpacs,final,nofootinbib]{revtex4}

\usepackage[polish,english]{babel}
\usepackage[latin1]{inputenc}
\usepackage{graphicx}
\usepackage{times}
\usepackage{bbold}
\usepackage[T1]{fontenc}
\usepackage{epsfig}
\usepackage{epstopdf}
\usepackage{bbold}
\usepackage{comment}
\usepackage{bm}

\usepackage{amsmath,amssymb}

\def\br{\begin{eqnarray}}
\def\er{\end{eqnarray}}
\def\be{\begin{equation}}
\def\ee{\end{equation}}

\def\({\left(}
\def\){\right)}

\def\rlx{\relax\leavevmode}
\def\IR{\rlx\hbox{\rm I\kern-.18em R}}

\def\vt{\vartheta}
\def\u2{\mid u\mid^2}

\newcommand{\gdhi}{\ooalign{\hfil/\hfil\crcr$\partial$}}
\newcommand{\gdAi}{\ooalign{\hfil/\hfil\crcr$A$}}

\newcommand{\sbr}[2]{\left\lbrack\,{#1}\, ,\,{#2}\,\right\rbrack}

\begin{document}

\title{Collective coordinate quantization and spin statistics of the solitons in the $\mathbb{C}P^N$ Skyrme-Faddeev model}
\author{Yuki~Amari$^{a}$}
\email{amalyamary@gmail.com}
\author{Pawe\l~Klimas$^{b}$}
\email{pawel.klimas@ufsc.br}
\author{Nobuyuki Sawado$^{a}$}
\email{sawado@ph.noda.tus.ac.jp}

\vspace{.5 in}
\small

\affiliation{
$^a$ Department of Physics, Tokyo University of Science,\\
 Noda, Chiba 278-8510, Japan \\
$^{b}$ Universidade Federal de Santa Catarina, Trindade, 88040-900, Florian\'opolis, SC, Brazil
}

\date{\today}

\begin{abstract}
The $\mathbb{C}P^N$ extended Skyrme-Faddeev model possesses planar soliton solutions. 
We consider quantum aspects of the solutions applying collective coordinate quantization 
in regime of rigid body approximation.  In order to discuss statistical properties of 
the solutions we include an Abelian Chern-Simons term (the Hopf term) in the Lagrangian. 
Since $\Pi_3(\mathbb{C}P^1)=\mathbb{Z}$ then for $N=1$ the term becomes an integer. 
On the other hand for $N>1$ it became perturbative  because $\Pi_3(\mathbb{C}P^N)$ is trivial. 
The prefactor of the Hopf term 
(anyon angle) $\Theta$ is not quantized and its value depends on the physical system. 
The corresponding fermionic models can fix value of the angle $\Theta$ for all $N$ in a way that the soliton with $N=1$ is not an anyon type
whereas for $N>1$ it is always an anyon even for $\Theta=n\pi, n\in \mathbb{Z}$. 
We quantize the solutions and calculate several mass spectra for $N=2$. 
Finally we discuss generalization for $N\geqq 3$.  
 \end{abstract}

\pacs{11.27.+d, 11.10.Lm, 11.30.-j, 12.39.Dc}

\maketitle 

\section{Introduction}

The Skyrme-Faddeev model is an example of a field theory that supports the finite-energy knotted solitons~\cite{sf}. 
Similarly to many other models \cite{coleman} the classical soliton solutions of the Skyrme-Faddeev model 
can play a role of adequate normal models useful in description of the strong coupling sector of the Yang-Mills theory. 
The exact soliton (vortex) solution of the model has been found within the integrable sector \cite{vortexlaf}.
The model contains some new quartic terms different to the Skyrme term. Inclusion of such terms is motivated by results of the analysis of the Wilsonian action of the $SU(2)$ Yang-Mills theory \cite{gies}.  
It has been shown that  in the case of the complex projective target space $\mathbb{C}P^N$ 
the extended Skyrme-Faddeev model possesses an exact soliton solution in the integrable sector provided that the coupling constants satisfy a special relation \cite{fk,fkz}. 
The existence of solutions of the model outside the integrable sector has been  confirmed numerically for appropriate choice of potentials~\cite{Amari:2015sva}. 

The research of quantum properties of solitons is important not only from a 
mathematical but also from a phenomenological point of view (mass spectrum, spin-statistics relation). 
There are many attempts to find a quantum theory of skyrmions in 2+1 dimensional $\mathbb{C}P^N$ model, 
including full canonical quantization scheme
 \cite{Haldane:1983ru,Bowick:1985ua,Kovner:1989wd,rodriquez,Marino:1999wg,Vlasii:2012kw,Vlasii:2014xda,Acus:2009df}. 
In this paper we shall generalize a scheme of quantization, usually discussed for $N=1$, to  an arbitrary value $N$. 
We begin our considerations presenting collective coordinate quantization of rotational degrees of freedom. 
The results could have some application to condensed matter physics, specifically,  to improve our comprehension of 
such phenomena as the nature of high $T_c$ superconductivity and also the fractional quantum Hall effect
~\cite{Haldane:1983ru,Bowick:1985ua,Kovner:1989wd,rodriquez,Marino:1999wg,Vlasii:2012kw,Vlasii:2014xda}.
There are already some important studies on a collective coordinate quantization approach to the Skyrme model with nonrelativistic
~\cite{Adkins:1983ya,Braaten:1988cc,Irwin:1998bs,Yabu:1987hm} and relativistic treatment~\cite{Hata:2010zy} which have as a 
goal an explanation of some basic properties of hadrons. A similar approach has been  applied to the Skyrme-Faddeev Hopfions~\cite{Krusch:2005bn,Kondo:2006sa}. 
An alternative approach based on a canonical quantization method has been already examined for the baby Skyrme model~\cite{Acus:2009df} and the Skyrme-Faddeev Hopfions~\cite{Acus:2012st}.

The Skyrme-Faddeev model on the $\mathbb{C}P^N$ target space in 3+1 dimensions is 
defined by the Lagrangian~\cite{fk}
\begin{eqnarray}
{\cal L}= -\frac{M^2}{2}{\rm Tr}\(\Psi^{-1}\partial_{\mu}\Psi\)^2
+\frac{1}{e^2}{\rm Tr}\(\sbr{\Psi^{-1}\partial_{\mu}\Psi}{\Psi^{-1}\partial_{\nu}\Psi}\)^2 \notag\\
+
\frac{\beta}{2}\left[{\rm Tr}\(\Psi^{-1}\partial_{\mu}\Psi\)^2\right]^2
+\gamma\left[{\rm Tr}\(\Psi^{-1}\partial_{\mu}\Psi\;
\Psi^{-1}\partial_{\nu}\Psi\)\right]^2-\mu^2V,
\label{actionp}
\end{eqnarray}
where $M^2$ is a coupling constant with dimension of square of mass whereas the coupling constants $e^{-2}$, $\beta$, $\gamma$ are dimensionless. 
The field $\Psi$ is called a principal variable
and it was extensively studied in \cite{fk} and also \cite{Amari:2015sva}. 
The Lagrangian is invariant under global transformation $\Psi\to {\cal A}\Psi {\cal B}^\dagger$
where ${\cal A},{\cal B}$ are some unitary matrices.  It turns out  that the zero modes of 
$\Psi$ impose an additional condition on matrices ${\cal A}$ and ${\cal B}$, namely, the asymptotic values  
of $\Psi$ must be preserved under the symmetry transformation, {\it i.e.},
${\cal A}\Psi_\infty {\cal B}^\dagger = \Psi_\infty$. 
There is no straightforward procedure how to obtain a 
suitable parametrization of the zero modes for $({\cal A}, {\cal B})$, however, one cannot exclude that such a parametrization exists.

In this paper, we shall deal with a slightly different parametrization of the field variable $\Psi$, namely, with the Hermitian variable $X$ obtained as a result of transformation $X:=C \Psi$ where $C$ is a dia\-gonal constant matrix $C:={\rm diag}(1,\cdots,1,-1)$. Note that such a transformation is a symmetry of the Lagrangian so one gets
\begin{align}
{\cal L}=\frac{M^2}{2}{\rm Tr}(\partial_\mu X\partial^\mu X)+\cdots\,.
\end{align}
The main advantage of this transformation is substitution of the asymptotic condition for the collective quantization by the following one  
${\cal A}X_\infty {\cal A}^\dagger =X_\infty$ where ${\cal A}\in SU(N)\otimes U(1)$. It is much easier to find a suitable parametrization consistent for the new condition. Thus the standard method for the quantization developed in 
\cite{Adkins:1983ya,Braaten:1988cc,Irwin:1998bs,Yabu:1987hm,Hata:2010zy,Krusch:2005bn,Kondo:2006sa,Acus:2009df,Acus:2012st} can be directly applied 
to the Hermitian variable $X$, than the principal variable $\Psi$ itself.  
\footnote{The situation is somewhat similar with the 
case of the 3+1 Skyrme model and the 2+1 baby Skyrme model. The former possesses the symmetry $U\to AUB^\dagger$ while
the latter only has the diagonal ones. It originates in the fact that the chiral symmetry can only be defined for 
odd space dimensions.}.

It is widely known that quantum aspects of the soliton solutions exhibit a special property (``fractional'' spin-statistics) when 
the Hopf term (theta term) is included in the action of the model~\cite{Wilczek:1983cy}. 
Since $\Pi_3 (\mathbb{C}P^1)=\mathbb{Z}$, then such a term became the Hopf invariant and therefore it can be represented as a 
total derivative which has no influence on classical equations of motion~\cite{Wu:1984kd}. On the other hand, since $\Pi_4(\mathbb{C}P^1)$ is trivial, the coupling constant (prefactor) $\Theta$ is not quantized. 
As shown in~\cite{Wilczek:1983cy}, when the Hopf Lagrangian is included in the model, the solitons with unit topological charge acquire fractional spin $\frac{\Theta}{2\pi}$. 
For a fermionic model coupled with $\mathbb{C}P^N$ field, $\Theta$ can be determined at least perturbatively~\cite{Abanov:2000ea,Abanov:2001iz}.  

$\Pi_3(\mathbb{C}P^N)$ is trivial for $N>1$ and then the Hopf term is perturbative, {\it i.e.}, it is not a homotopy invariant. It means that the contribution from this term can  be fractional even for an integer $n$ in the anyon angle $\Theta=n\pi$.
It was pointed out in~\cite{Bar:2003ip} that  
an analogue of the Wess-Zumino-Witten term appears for the $\mathbb{C}P^N$ field and it plays a similar role as the Hopf term for $N=1$~\cite{Jaroszewicz:1985ip}. 
Consequently, the soliton can be quantized as an anyon with statistics angle $\Theta$ and also such Hopf-like term. 

The paper ~\cite{Kovner:1989wd} contains discussion of the influence of this term on quantum spectra  for $N>1$. The author has taken into account the field being a trivial extension of the case $N=1$ ({\it i.e.}, including  only a single winding number). 
In this paper we shall give more thorough and complete discussions of quantum spectra for $N>1$ implementing a set of winding numbers $n_1,n_2,\cdots$. 
We shall present the quantum spectra within a standard semiclassical zero mode quantization scheme. 

The paper is organized as follows. In Sec.II we give a brief review of the extended Skyrme-Faddeev model on the $\mathbb{C}P^N$ target space, its classical solutions and their topological charges.  The Hopf Lagrangian is presented in the final part of this section. 
In Sec.III we briefly discuss the quantization scheme and fractional spin of solitons in the model with $N=1$ (baby skyrmion). 
Section IV contains  generalization of the collective coordinate quantization scheme for the case $N=2$. In Sec.V we present the analysis of the spectrum. 
Finally in Sec.VI we generalize our formula for $N\geqq 3$ and we present the energy plot of the quantized system.
Section VII contains summary of the paper.


\section{The $\mathbb{C}P^N$ extended Skyrme-Faddeev model}
The extended Skyrme-Faddeev model on the $\mathbb{C}P^N$ target space
 has been proposed in \cite{fk}. The coset space $\mathbb{C}P^N=SU(N+1)/SU(N)\otimes U(1)$ is an example 
of a symmetric space and it can be naturally parametrized in terms of so called {\it principal variable} 
$\Psi(g):=g\sigma(g)^{-1}$, with $g\in SU(N+1)$, $\sigma$ being the order two automorphism under which the subgroup 
$SU(N)\otimes U(1)$ is invariant {\it i.e.} 
$\sigma(h)=h$ for $h\in SU(N)\otimes U(1)$. 
The principal coordinate $\Psi(g)$ defined above satisfies $\Psi(gh)=\Psi(g)$. 
Therefore we have just one matrix $\Psi(g)$ for each coset in $SU(N+1)/SU(N)\otimes U(1)$.

The first term of the Lagrangian (\ref{actionp}) is quadratic in $\Psi$ and corresponds with the Lagrangian of the $\mathbb{C}P^N$ model. 
The quartic term proportional to $e^{-2}$ is the Skyrme term whereas other quartic terms constitute the extension of standard Skyrme-Faddeev model. 
The non-Skyrme type  quartic terms introduce to the Lagrangian some fourth power time derivative terms. A form of the Lagrangian adequate for quantization is obtained imposing a condition
\begin{eqnarray}
\beta+2\gamma=0
\label{properhamiltonian}
\end{eqnarray}
which eliminates some unwanted terms.  
We shall analyze in this paper some solutions of the 2+1 dimensional model (a planar case). 
In such a case the coupling constants have different physical dimensions to those in the 3+1 dimensional model, {\it i.e.},  
$M$ has dimension of mass$^{1/2}$ and three other coupling constants $e^{-2},\beta,\gamma$ have dimension of mass.

According to the previous paper \cite{fk}, one can parametrize the model in terms of $N$ 
complex fields $u_i$, where $i=1,...,N$. We assume an $(N+1)$-dimensional defining 
representation where the $SU(N+1)$ valued element $g$ is of the form 
\begin{eqnarray}
	g\equiv\frac{1}{\vartheta}\left( \begin{array}{cc}
	\Delta & iu \\ 
	iu^\dagger & 1
	\end{array} \right) 
	\qquad
	\vartheta\equiv\sqrt{1+u^\dagger\cdot u}
\end{eqnarray}  
and where $\Delta$ is the Hermitian $N\times N$-matrix, 
\begin{equation}
	\Delta_{ij}=\vartheta\delta_{ij}-\dfrac{u_iu_j^*}{1+\vartheta}
	~~\text{which satisfies}~~
	\Delta\cdot u=u
	~\text{and}~
	 u^\dagger\cdot\Delta=u^\dagger.
	 \notag
\end{equation}
The principal variable takes the form
\begin{equation}
	\Psi(g)=g^2=	\(\begin{array}{cc}
	I_{N\times N} & 0 \\
	0 & -1 
	\end{array}\)+
	\frac{2}{\vt^2}\(\begin{array}{cc}
	-u\otimes u^\dagger & iu \\
	iu^\dagger & 1  
	\end{array}\).
\end{equation}
It has been shown recently that the model (\ref{actionp}) possesses vortex solutions. 
There exists a family of exact solutions in the model without potentials where in addition the coupling constants 
satisfy the condition $\beta e^2+\gamma e^2=2$. The solutions satisfy the zero curvature condition 
$\partial_\mu u_i \partial^\mu u_j=0$ for all $i,j=1,...,N$ and therefore one can construct the infinite set of conserved currents.
Furthermore, according to numerical study there exist vortex solutions which do not belong to the integrable sector. 
Such solutions have been found for the potential 
\begin{eqnarray}
V={\rm Tr}(1-{\Psi_{0}}^{-1}\Psi)^a{\rm Tr}(1-{\Psi_{\infty}}^{-1}\Psi)^b  
\label{potcp02}
\end{eqnarray}
with $a\geq 0, b>0$ where $\Psi_0$ and $\Psi_{\infty}$ are a vacuum value of the field 
$\Psi$ at origin and spatial infinity respectively. The potential (\ref{potcp02}) is an analog of potentials for the baby Skyrme model.
The numerical solutions and holomorphic exact solutions corresponding with the same set of winding numbers have common boundary behavior. 

The Lagrangian (\ref{actionp}) is invariant under the global symmetry $\Psi\to {\cal A}\Psi {\cal B}^\dagger, {\cal A, B}\in SU(N+1)$. 
Since we restrict the analysis to 2+1 dimensions then
it is natural to consider a diagonal subgroup. For this reason we transform the variable $\Psi$ into the Hermitian one
\begin{eqnarray}
X:=
I_{N+1\times N+1}
+
\frac{2}{\vt^2}\(\begin{array}{cc}
-u\otimes u^\dagger & iu \\
-iu^\dagger & -1  
\end{array}\) \label{HermitianX}
\end{eqnarray}
which in addition satisfies $X^{-1}=X$. 
Now the Lagrangian (\ref{actionp}) becomes
\begin{align}
&{\cal L}=\frac{M^2}{2}{\rm Tr}\(\partial_{\mu}X\partial^\mu X\)
+\frac{1}{e^2}{\rm Tr}\(\sbr{\partial_{\mu}X}{\partial_{\nu}X}\)^2 \notag\\
&+
\frac{\beta}{2}\left[{\rm Tr}\(\partial_{\mu}X\partial^\mu X\)\right]^2
+\gamma\left[{\rm Tr}\(\partial_{\mu}X\;
\partial_{\nu}X\)\right]^2-\mu^2V(X)\,.
\label{actionx}
\end{align}
An analysis of zero modes of the classical solutions is much easier in approach involving a variable $X$ and in practice enables to apply the quantization scheme. In order to explain this statement let us note that for the variable $\Psi$ in the Lagrangian (\ref{actionp}) the boundary conditions which result in ${\cal A}\Psi_\infty {\cal B}^\dagger=\Psi_\infty$ break partially the symmetry  associated with the transformation $\Psi\to {\cal A}\Psi {\cal B}^\dagger$.
Unlike for the standard skyrmion, where the chiral field $U$ goes to $U_\infty=I$ 
and the symmetry is simply broken down to ${\cal A}={\cal B}$, the $\Psi_\infty$ has a nontrivial value which depends on winding numbers. Moreover, one still has to determine the pair of $({\cal A}, {\cal B})$ for the zero-modes. 
On the contrary, for (\ref{actionx}) the symmetry transformation becomes diagonal, {\it i.e.}, $X\to {\cal A}X{\cal A}^\dagger$, and  then an explicit form of ${\cal A}$ can be easily determined as expansion in basis of the standard Gell-Mann matrices.  

It is worth it to  stress that for the planar case the classical equations of motion and their classical solutions  have exactly the same form for both parametrizations. Furthermore, since the quantization procedure is based on properties of classical solutions then the resulting quantum spectra must correspond. 

The variable $X$ has close relation with a well-known Hermitian projector $P$ that satisfies
\begin{eqnarray}
P^\dagger=P,~~{\rm Tr}P=1,~~P^2=P\,.
\end{eqnarray}
The projector $P$ is defined as
\begin{eqnarray}
P(V)={\cal Z}\otimes {\cal Z}^\dagger
\end{eqnarray}
where the symbol ${\cal Z}$ stands for the $N$-component complex vector
${\cal Z}=(u_1,\ldots,u_N,i)^T/\sqrt{1+u^{\dagger}\cdot u}$ which depends on two variables $z,z^*$. The form of the projector allows to express $X$ in the form
\begin{eqnarray}
X=I_{N+1\times N+1}-2P\,.
\label{relationofxp}
\end{eqnarray}

We introduce dimensionless coordinates $(t,\rho,\varphi)$ 
\br
x^0=r_0t,\quad x^1=r_0\rho\cos\varphi,\quad x^2=r_0\rho\sin\varphi
\er
where the length scale $r_0$ is defined in terms of coupling constants $M^2>0$ and $e^2<0$  {\it i.e.}
$
r_0^2:=-\dfrac{4}{M^2e^2}
$
and the light speed is $c=1$ in the natural units. The linear element $ds^2$ reads
$$
ds^2=r_0^2(dt^2-d\rho^2-\rho^2d\varphi^2).
$$
We shall consider the axial symmetric planar solutions 
\br
u_j=f_j(\rho)e^{in_j\varphi}\label{ansatz}
\er
where the constants $n_i$ form the set of integer numbers and $f_i(\rho)$ are real-valued functions.
Equivalently, the ansatz (\ref{ansatz}) in matrix form reads 
$u=f(\rho)e^{i\lambda\varphi}$ where $\lambda={\rm diag}(n_1,\ldots,n_N)$.
In order to simplify the form of some formulas below, we introduce the functions defined as follows
\begin{equation}
\begin{split}
\theta&=-\frac{4}{\vt^4}\,\left[\vt^2\,f'^T.f'-(f'^T.f)(f^T.f')\right],\\
\omega&=-\frac{4}{\vt^4}\,\left[\vt^2\,f^T.\lambda^2.f-(f^T.\lambda.f)^2\right],\\
\zeta&=-\frac{4}{\vt^4}\,\left[\vt^2\,f'^T.\lambda.f-(f^T.\lambda.f)(f'^T.f)\right] 
\end{split}
\end{equation}
where derivative with respect to $\rho$ is denoted by $\frac{d}{d\rho}='$ and $T$ stands for matrix transposition. The classical equations of motion written in dimensionless coordinates take the form
\begin{align}
&(1+f^T.f)\left[\frac{1}{\rho}\left(\rho\,C_{1}f'_k\right)'+\frac{i}{\rho}\left(\frac{C_{3}}{\rho}\right)'(\lambda.f)_k-\frac{1}{\rho^4}C_{2}(\lambda^2.f)_k\right] \notag\\
&-2\left[C_{1}(f^T.f')f'_k-\frac{1}{\rho^4}C_{2}(f^T.\lambda.f)(\lambda.f)_k\right]\nonumber\\
&+\tilde{\mu}^2\frac{f_k}{4}(1+f^{T}.f)^2\left[\frac{\delta V}{\delta f_k^2}+\sum_{i=1}^Nf_i^2\frac{\delta V}{\delta f_i^2}\right]=0
\label{equationr}
\end{align}
for each $k=1,\ldots,N$, where 
$\tilde{\mu}^2:=\frac{r_0^2}{M^2}\mu^2$ and symbols $C_j$ take the form
\begin{equation}
\begin{split}
 &C_{1}=-1+(\beta e^2-1)\frac{\omega}{\rho^2},\\
 &C_{2}=-\rho^2+\rho^2(\beta e^2-1)\theta,\\
 &C_{3}=3i\zeta.
\end{split}
\end{equation}
The energy of the static solution is given by the integral
\begin{eqnarray}
 M_{\rm cl}=-2\pi M^2 \int \rho d\rho 
\left( \theta+\frac{\omega}{\rho^2}+\frac{3\zeta^2}{\rho^2}-(\beta e^2-1)\frac{\theta\omega}{\rho^2}
-\tilde{\mu}^2 V \right). \nonumber \\
 \label{ClassicalMass}
\end{eqnarray}

According to discussion in \cite{D'Adda:1978uc} and also in \cite{fkz}
one can introduce two-dimensional topological charges associated with vortex configurations. Such charges are closely related with a topological current that  has the following form in terms of  the principal variable
\begin{eqnarray}
j^\mu (X)=\frac{i}{16\pi}\epsilon^{\mu\nu\lambda}{\rm Tr}(X\partial_\nu X\partial_\lambda X).
\label{topologicalcurrent}
\end{eqnarray}
Since the solutions behave as holomorphic functions near the boundaries then the topological charges are equal 
to the number of poles of $u_i$, including those at infinity, {\it i.e.},
\begin{equation}
Q_{\rm top}=\int j^0(X)d^2x=n_{\rm max}+|n_{\rm min}|
\label{topologicalcharge}
\end{equation}
where $n_{\rm max}$ is the highest positive integer in the set $n_i, i=1,2,\cdots,N$ and $n_{\rm min}$ is the lowest negative integer in the same set. 

The conserved current (\ref{topologicalcurrent}) defines a gauge potential 
\begin{eqnarray}
j^\mu=-\frac{i}{2\pi}\epsilon^{\mu\nu\lambda}\partial_\nu a_\lambda
\end{eqnarray}
where $a_\mu$ is determined up to the gauge freedom $a_\mu\to a_\mu-\partial_\mu\Lambda$. 
As it was pointed out in \cite{Wilczek:1983cy}, $a_\mu$ is a nonlocal function of $X$.
The straightforward calculation shows that $a_\mu$ can be written as 
\begin{eqnarray}
a_\mu=-2\pi i\partial^{-2}[\epsilon_{\mu\nu\lambda}\partial^\nu j^\lambda],~~~~~
{\rm in~the~gauge}~~\partial^\mu a_\mu=0.
\end{eqnarray}
In the alternative approach the gauge potential $a_\mu$ is given in terms of the complex vector ${\cal Z}$ 
\begin{eqnarray}
a_\mu=-i{\cal Z}^\dagger\partial_\mu{\cal Z}
\end{eqnarray}
where the U(1) rotation acting on ${\cal Z}$ induces the gauge transformation on $a_\mu$.

The ``Hopf Lagrangian'' is defined in terms of $a_{\mu}$ and it reads
\begin{eqnarray}
\Theta{\cal L}_{\rm Hopf}=-\frac{\Theta}{4\pi^2}\epsilon^{\mu\nu\lambda}a_\mu\partial_\nu a_\lambda.
\label{action_hopf}
\end{eqnarray}
This Lagrangian is invariant under U(1) gauge transformation and the value of the prefactor $\Theta$ is essentially undetermined.
Since $\Pi_3 (\mathbb{C}P^1)=\mathbb{Z}$ then (\ref{action_hopf}) is exactly the Hopf invariant for $N=1$ and consequently it can be expressed as a total derivative.  For this reason it does not contribute to the classical equations of motion. 
On the contrary, $\Pi_3(\mathbb{C}P^N)$ is trivial for $N>1$  and therefore the Hopf term is 
not a homotopic invariant in this case. It means that the contribution from the Hopf term is always fractional even for an integer $m$ in the anyon angle $\Theta=m\pi$. 
Note that even though the Hopf term is not a total derivative anymore, it still does not affect the classical soliton solutions because it is linear in time derivative. 

In the following part we quantize the model containing the Lagrangian (\ref{actionx}) extended by the Hopf term (\ref{action_hopf}) and examine the spin statistics of the $\mathbb{C}P^N$ solitons.

\section{Collective coordinate quantization of the baby skyrmions}

It became quite instructive to present a scheme of quantization for the model with the 
$\mathbb{C}P^1$ target space before going to the main question which is a quantization 
of the model with the  $\mathbb{C}P^N$ target space. The model with $N>1$ is 
technically more complex because it contains many fields.
For this reason we shall begin presenting analysis of the  $\mathbb{C}P^1$  
baby skyrmions. The full canonical quantization of the model has already been
studied \cite{Acus:2009df}, however, in absence of the Hopf term. 
We consider the collective coordinate quantization taking into account the 
Hopf term and discuss the spin of the baby skyrmions.
 
\subsection{The model and the quantized energy}

The baby Skyrme model~\cite{Piette:1994ug,Kudryavtsev:1996er} is a mimic of 
a hadronic Skyrme model. Its solutions (baby skyrmions) are considered as possible 
candidates for vortices or spin textures.

The  model is given in terms of a vectorial triplet $\vec{n}=(n_1,n_2,n_3)$ with the 
constraint $\vec{n}\cdot\vec{n}=1$.  Performing stereographic projection $S^2$
on a complex plane one can parametrize the model by a complex scalar field $u$ related to 
the triplet $\vec{n}$ by formula
\begin{eqnarray}
\vec{n}=\frac{1}{1+|u|^2}(u+u^*,-i(u-u^*),|u|^2-1). 
\label{stereographic}
\end{eqnarray}
Instead, we shall make use of another alternative parametrization that 
is convenient for any $\mathbb{C}P^{N}$ space, in particular, 
also for $SU(2)/U(1)=\mathbb{C}P^1$  coset space. In such a case the Hermitian principal
variable (\ref{HermitianX}) $X$ is a function of just one complex field $u$ 
\begin{eqnarray}
X=\frac{1}{1+|u|^2}
\(\begin{array}{cc}
1-|u|^2 & 2iu \\
-2iu^* & |u|^2 -1 
\end{array}\).
\label{CP1 principal}
\end{eqnarray}
It can be also expressed in terms of components of the unit vector $\vec{n}$ 
\begin{eqnarray}
X=-n^3\tau_3-n^2 \tau_1-n^1\tau_2.
\end{eqnarray}

The Lagrangian of the baby Skyrme model parametrized by the variable $X$ takes the form  
\begin{eqnarray}
{\cal L}_{\rm bS}=\frac{M^2}{2}{\rm Tr}(\partial_\mu X\partial^\mu X)
-\frac{1}{8e^2}{\rm Tr}([\partial_\mu X,\partial_\nu X]^2)
-\mu^2V
\label{action_bs}
\end{eqnarray}
where $M^2,e^2$ are coupling constant of the model and $V$ is a
potential which we shall not specify for a moment because its explicit form is irrelevant for  
current discussion. 
We shall consider a model ${\cal L}:={\cal L}_{\rm bS}+\Theta{\cal L}_{\rm Hopf}$ which 
constitutes extension of the model (\ref{action_bs}) due to the Hopf term (\ref{action_hopf}).
Since $\Pi_3 (\mathbb{C}P^1)=\mathbb{Z}$, the Hopf term can be represented as a 
total derivative so it does not  contribute to the classical equations of motion. 
The complex coordinate $Z$ and the Hermitian principal variable $X$ in the 
case of the $\mathbb{C}P^1$ target space are related as
$X:=1-2Z\otimes Z^\dagger$. The topological charge is given by 
\begin{eqnarray}
q_{\rm top}=\frac{i}{16\pi}\int d^2x\epsilon_{ij}{\rm Tr}(X\partial_iX\partial_jX),~~
i,j=1,2\,.
\label{charge_bs}
\end{eqnarray}

Note that expressions (\ref{action_bs}) and (\ref{charge_bs}) are invariant under rotation 
realized by a unitary matrix $A$ according to  transformation $X\rightarrow A X A^\dagger$.
The standard procedure proceeds by promoting the parameter $A$ to 
the status of dynamical variable $A(x_0)$. Then the dynamical 
ansatz adopted in collective coordinate quantization reads
\br
X(\bm{r};A(x_0))=A(x_0)X(\bm{r})A^\dagger(x_0).
\label{DynamicalAnsatz_bs}
\er
The expression (\ref{DynamicalAnsatz_bs}) parametrized by a complex coordinate $Z$ reads
\begin{eqnarray}
X(\bm{r},A(x_0))=A(x_0)(1-2Z(\bm{r})Z^\dagger (\bm{r}) )A^\dagger (x_0) \nonumber \\
=1-2\bigl(A(x_0)Z(\bm{r})\bigr)\bigl(A(x_0)Z(\bm{r})\bigr)^\dagger
\end{eqnarray}
which allows us to conclude that
\begin{eqnarray}
Z(\bm{r},A(x_0))=A(x_0)Z(\bm{r})\,.
\label{DynamicalAnsatzZ_bs}
\end{eqnarray}
Plugging (\ref{DynamicalAnsatz_bs}) into the Lagrangian (\ref{action_bs}) and also
(\ref{DynamicalAnsatzZ_bs}) into the Hopf term (\ref{action_hopf}) we obtain 
an effective Lagrangian
\begin{equation}
L_{\rm eff}=\frac{1}{2}I_{ab}\Omega_a\Omega_b +\frac{\Theta}{4\pi}\Lambda_a\Omega_a-M_{\rm cl}
\label{effectiveL1}
\end{equation}
where the collective angular velocities $\Omega_a$ appear in expansion of
 the operator $iA^\dagger\partial_{x_0}A=\frac{\tau_a}{2}\Omega_a$ and where $\tau_a$ are Pauli matrices $a=1,2,3$. 
The inertia tensor $I_{ab}$ is given in terms of $X(\bm{r})$ 
\begin{eqnarray}
 I_{ab}&=&-\frac{4}{e^2} \int\rho d\rho d\varphi
\biggl\{
-{\rm Tr}\Bigl(\bigl[\frac{\tau_a}{2},X\bigr]\bigl[\frac{\tau_b}{2},X\bigr]\Bigr) \nonumber \\
&+&\frac{1}{8}{\rm Tr}\Bigl(\Bigl[\bigl[\frac{\tau_a}{2},X\bigr],\partial_k X\Bigr]
\Bigl[\bigl[\frac{\tau_b}{2},X\bigr],\partial_k X\Bigr]\Bigr)
\biggr\}.
\label{InertiaTensor1}
\end{eqnarray}
We consider the well known ``hedgehog'' ansatz
\begin{eqnarray}
&\vec{n}=(\sin g(\rho)\cos n\varphi,\sin g(\rho)\sin n\varphi,\cos g(\rho)) \nonumber \\
&g(0)=\pi,~ g(\infty )=0.0.
\label{hedgehog ansatz}
\end{eqnarray}
obtained from (\ref{stereographic}) by the following parametrization   of the complex field $u$
\begin{equation}
	u=\cot\frac{g(\rho)}{2}e^{in\varphi}.
\end{equation}
The topological charge of solutions \eqref{hedgehog ansatz} takes integer values according to
\begin{eqnarray}
q_{\rm top}=-\frac{n}{2}\int_0^\infty d\rho \sin gg'=n.
\end{eqnarray}
The components of the moment of inertia read
\begin{eqnarray}
&&I_{11}=I_{22} \nonumber \\
&&~~~~=-\frac{4\pi}{e^2}\int^\infty_0\rho d\rho
\biggl(2+2\cos^2g+\frac{n^2\sin^2g}{\rho^2}+\cos^2gg'^2\biggr) \nonumber \\ \\
&&I_{33}=-\frac{16\pi}{e^2}\int^\infty_0\rho d\rho \sin^2g(1+2g'^2)
\end{eqnarray}
and $I_{ab}=0$ for $a\neq b$. 
Note that  rotation is allowed only around the third axis because $I_{11}=I_{22}=\infty$
~\footnote{Our results are essentially equivalent with (19) of~Ref.\cite{Acus:2009df}
up to constant. However, their formulation (20) (corresponding to our (37)) was incorrect about sign of the coefficient.}.

Taking into account that $\Pi_3(\mathbb{C}P^1)=\mathbb{Z}$ we obtain integer values for  
expressions $\Lambda_a$ that appear in the Hopf term
\begin{eqnarray}
\Lambda_3&=&-i\int d\rho \biggl((Z^\dagger \tau^3Z)\partial_\rho (Z^\dagger\partial_\varphi Z)
-(Z^\dagger\partial_\varphi Z)\partial_\rho (Z^\dagger \tau^3Z)\biggr) \nonumber \\
&=&-n
\label{lambda3_bs}
\end{eqnarray}
and $\Lambda_1=\Lambda_2=0$. The (body-fixed) isospin operator $J_3$ can be introduced as a symmetry transformation generator via  Noether's theorem  
\begin{eqnarray}
A\to Ae^{i\frac{\tau_3}{2}\vartheta_3}, \nonumber \\
J_3=-I_{33}\Omega_3+\frac{n\Theta}{4\pi}\,.
\end{eqnarray}
The Legendre transform of the Lagrangian leads to the following expression for the Hamiltonian
\begin{eqnarray}
H_{\rm eff}=M_{\rm cl}+\frac{g_{33}}{2}\biggl(J_3-\frac{n\Theta}{4\pi}\biggr)^2\,.
\end{eqnarray}
where $g_{33}$ is inverse of the moments of inertia $I_{33}$ {\it i.e.} $g_{33}:=1/I_{33}$.
If one represents the isospin operator $J_3:=i\frac{\partial}{\partial\alpha}$ as acting on 
the basis $|\ell\rangle \equiv e^{-ik\alpha}|0\rangle $,
with $k$ being an integer or a half-integer numbers, then
the energy eigenvalue is given by the expression 
\begin{eqnarray}
E=M_{\rm  cl}+\frac{g_{33}}{2}\biggl(k-\frac{n\Theta}{4\pi}\biggr)^2\,.
\end{eqnarray}
One can substitute  the quantum number $k$ by an integer-valued index {\it i.e.}, $\ell\equiv 2k$ what gives
\begin{eqnarray}
E=M_{\rm  cl}+\frac{\tilde{g}_{33}}{2}\biggl(\ell-\frac{n\Theta}{2\pi}\biggr)^2
,~~~~\ell\in \mathbb{Z}
\end{eqnarray}
where $\tilde{g}_{33}=g_{33}/4$. This is a familiar result: for $n\Theta=0$ or in general 
({\it even} number)$\times \pi$, the angular momentum is integer then one gets boson, while for 
$n\Theta=\pi$ or ({\it odd} number) $\times \pi$, the angular momentum is half
integer then one gets fermion. 

\subsection{The fermionic effective model and the anyon angle}

It is well known that for a fermionic effective model coupled to 
a baby skyrmion with a constant gap $m$ the integrating out the Dirac 
field leads to effective Lagrangian containing a 
kind of baby Skyrme model and some topological terms including 
the Hopf term~\cite{Abanov:2000ea,Abanov:2001iz}. 
The Euclidean path integral of the partition function, which enables us to 
examine the topological term after integrating out the Dirac field, is of the form
\begin{eqnarray}
\Gamma (X,A_\mu)=\int {\cal D}\psi{\cal D}\bar{\psi} 
\exp \biggl(\int d^3x \bar{\psi}iD\psi\biggr)
\label{partition function}
\end{eqnarray}
where the U(1) gauged Dirac operator reads 
\begin{eqnarray}
iD:=i\gamma^\mu (\partial_\mu-iA_\mu)-mX.
\end{eqnarray} 
A number of articles extensively describe the derivative expansion of 
the effective action $S_{\rm eff}$ that appears in $\Gamma:=\exp(S_{\rm eff})$. 
It contains both the action of the model (in the real part) and the topological terms
 (in the imaginary part). 
After a bit lengthy calculation (see Appendix A) one gets
\begin{eqnarray}
&&{\rm Re}S_{\rm eff}=\frac{|m|}{2\pi}\int d^3x{\rm Tr}(\partial_\mu X\partial^{\mu} X)
+O(\partial X^3),
\label{effectiveaction_fr}
\\
&&{\rm Im}S_{\rm eff}=\int d^3x\Bigl(j^\mu (X) A_\mu-\pi\,{\rm sgn}(m){\cal L}_{\rm Hopf}(X)\Bigr)\,.
\label{effectiveaction_fi}
\end{eqnarray}
The explicit form of the current $j^\mu$ coincides with (\ref{topologicalcurrent}).
Consequently, as pointed out in \cite{Jaroszewicz:1985ip,Abanov:2000ea,Abanov:2001iz},
the anyon angle $\Theta$ is determinable in this fermionic context. 
It means that  the soliton became a fermion for odd topological charges and  a boson for even topological charges. 

\section{Collective coordinate quantization of the $\mathbb{C}P^2$ model}

It has been already mentioned that the Lagrangian  density (\ref{actionp}) is invariant under transformation
$\Psi\rightarrow \mathcal{A}\Psi\mathcal{B}^\dagger$ where $\mathcal{A},\mathcal{B}$ are some unitary constant matrices. This symmetry could remain also for the Lagrangian density written in new variable $X=C\Psi$ where $C$ is a diagonal constant matrix. However, for $X=X^\dagger$ only the diagonal symmetry ${\cal A}={\cal B}$ is allowed.  
Moreover, the topological charge (\ref{topologicalcharge}) is invariant under such transformation 
only for $\mathcal{A}=\mathcal{B}$. It leads to the conclusion that for 
a model which supports topological soliton solutions the only allowed symmetry is a diagonal one $X\to{\mathcal A}X{\mathcal A}^\dagger$. 
In fact there is another restriction on $\mathcal{A}$, namely 
for the asymptotic field $X_\infty$ it must hold\begin{equation}
	\mathcal{A}X_\infty \mathcal{A}^\dagger=X_\infty,
	\label{ConditionInfty}
\end{equation}
otherwise the moments of inertia  corresponding to the modes diverge.
Note that expression ${\cal A}$ satisfying (\ref{ConditionInfty}) depends on  numbers $(n_1,n_2)$ because these numbers determine the form of $X_\infty$. 
The vortex solutions are symmetric under exchange of $n_1,n_2$.
It is enough to study configurations with $n_1>n_2$ where the cases $n_1>0,n_1<0$ are treated separately. 

In analogy to baby skyrmions we shall adopt following dynamical ansatz in collective coordinate quantization 
\br
X(\bm{r};\mathcal{A}(x_0))=\mathcal{A}(x_0)X(\bm{r})\mathcal{A}^\dagger(x_0).
\label{DynamicalAnsatz}
\er
Substituting (\ref{DynamicalAnsatz}) into the Lagrangian (\ref{actionx}) one obtains a Lagrangian which depends on the collective angular velocity operator $i\mathcal{A}^\dagger\partial_{x_0}\mathcal{A}$. Such an operator possesses expansion on the set of collective coordinates which appear in the resulting effective Hamiltonian.  

\begin{figure*}[t]
\includegraphics[width=14cm,clip]{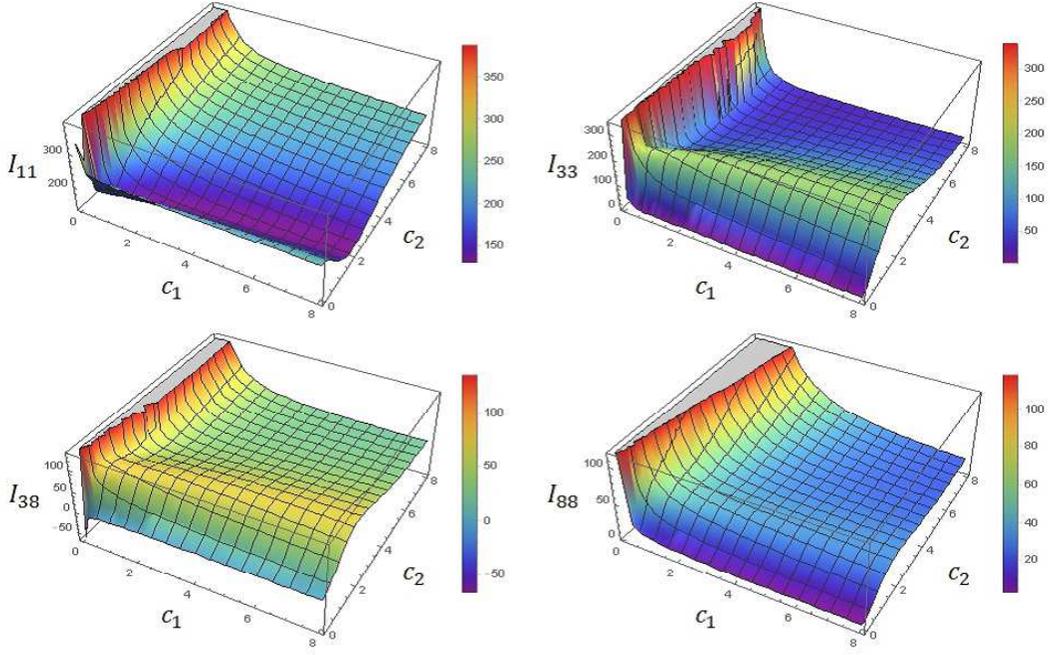}
\caption{\label{momentofinertia_holomorphic}The finite components of inertia tensor~(\ref{InertiaTensor}) 
of the holomorphic solutions for the topological charge $Q_{\rm top}=3$, i.e.,
$(n_1,n_2)=(3,1))$ in unit of $(-4/e^2)$.}
\end{figure*}

\begin{figure*}[t]
\includegraphics[width=14cm,clip]{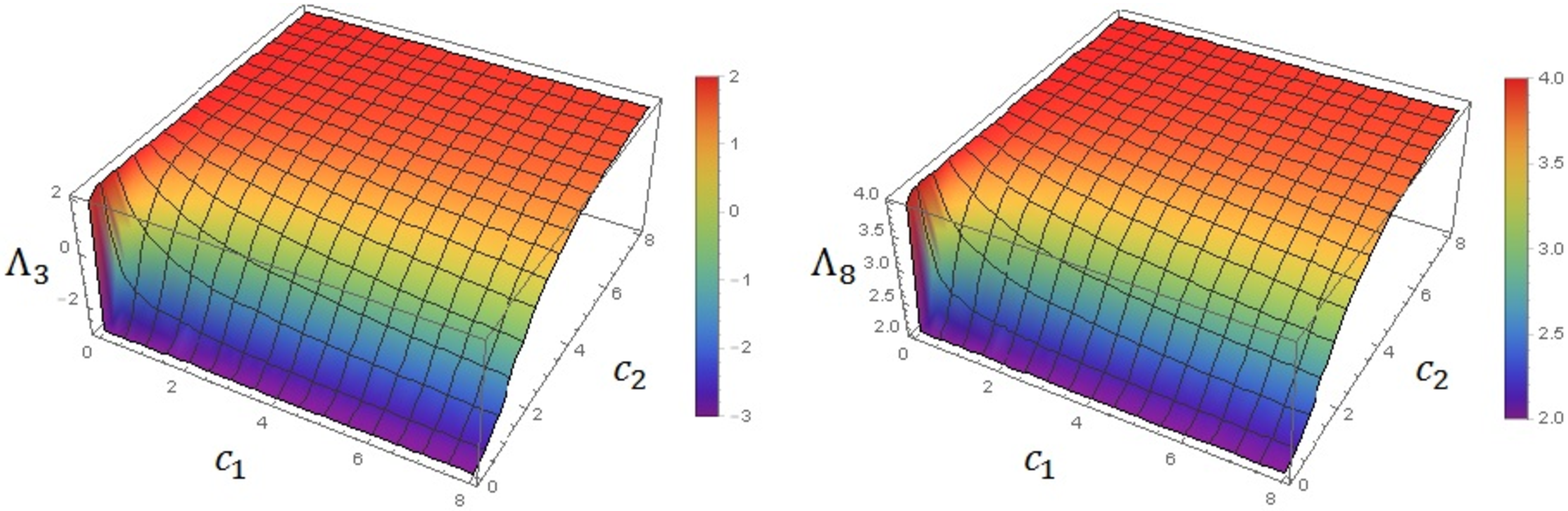}
\caption{\label{lambda_holomorphic}The finite components of inertia vector~(\ref{InertiaVector}) 
of the holomorphic solutions for the topological charge $Q_{\rm top}=3$, {\it i.e.},
$(n_1,n_2)=(3,1))$ in unit of $(-4/e^2)$.}
\end{figure*}

\subsection{The case $n_1>0$}
In this case the asymptotic value of the principal variable $X$ is $X_\infty=\mathrm{diag}(-1,1,1)$. 
The generators $\{F_a\},a=1,2,3,8$ of the symmetry (\ref{ConditionInfty}) have the form 
\begin{eqnarray}
&&F_1:=\frac{\lambda_6}{2},\qquad~ F_2:=\frac{\lambda_7}{2},\qquad~ F_3:=-\frac{1}{4}(\lambda_3-\sqrt{3}\lambda_8)
	\nonumber \\
&&\hspace{1.5cm}F_8:=-\frac{1}{2}\Bigl(\lambda_3+\frac{1}{\sqrt{3}}\lambda_8\Bigr)\,. \notag
\end{eqnarray}
They satisfy the commutation relations
\begin{equation}
	[F_a,F_b]=i\epsilon_{abc}F_c,  
	\quad [F_a,F_8]=0 \qquad a,b,c=1,2,3
\end{equation} 
what shows that the symmetry (\ref{ConditionInfty}) is in fact  a residual symmetry $SU(2)\times U(1)$. 
The rotation matrix $\mathcal{A}$ is parametrized by four Euler angles $\vartheta_i, (i=1,2,3,8)$ 
in the following way
\begin{equation}
\mathcal{A}=e^{-iF_3\vartheta_1}e^{-iF_2\vartheta_2}
e^{-iF_3\vartheta_3}e^{-iF_8\vartheta_8}.
\end{equation}
The angular velocities $\Omega_a$ of the collective coordinates became the expansion coefficients of 
the operator $i\mathcal{A}^\dagger\partial_{x_0}\mathcal{A}$ in a basis of generators $F_a$ of the residual symmetry. The expansion takes the form
 \begin{equation}
 	i\mathcal{A}^\dagger\partial_{x_0}\mathcal{A}=F_a\Omega_a.
 \end{equation}
  
The effective Lagrangian contains a term quadratic in $\Omega_a$ which comes from the 
Skyrme-Faddeev part of the total Lagrangian and a term linear in $\Omega_a$ having origin in the Hopf Lagrangian 
\begin{equation}
L_{\rm eff}=\frac{1}{2}I_{ab}\Omega_a\Omega_b +\frac{\Theta}{4\pi}\Lambda_a\Omega_a-M_{\rm cl}.
\label{effectiveL}
\end{equation}
where  the symmetric inertia tensor $I_{ab}$  is given as the integral of the expressions containing 
the Hermitian principal variable $X(\bm{r})$ and they read
\begin{align}
 I_{ab}=\frac{4}{e^2}&\int\rho d\rho d\varphi\biggl[
\mathrm{Tr}\Bigl(\left[F_a,X\right]\left[F_b,X\right]\Bigr)  \notag\\ 
&+\mathrm{Tr}\Bigl(\left[\left[F_a,X\right],\partial_kX\right]\left[\left[F_b,X\right],\partial_kX\right]\Bigr) \notag\\
&+\frac{\beta e^2}{2}\bigg\{ \mathrm{Tr}\Bigl([F_a,X] [F_b,X]\Bigr)
\mathrm{Tr}(\partial_kX\partial_kX) \notag\\ 
&\hspace{1cm}-\mathrm{Tr}\Bigl([F_a,X]\partial_kX\Bigr)\mathrm{Tr}\Bigl([F_b,X]\partial_kX\Bigr)
\bigg\}
\biggr].
\label{InertiaTensor}
\end{align}
The symmetry of components $I_{38}=I_{83}$ and equality $I_{11}=I_{22}$ originate in the axial symmetry imposed in the ansatz (\ref{ansatz}).

In generality the  Hopf term in the Lagrangian (\ref{effectiveL}) 
is nonlocal in fields $X$. However, if we translate the field into ${\cal Z}$ 
using transformation (\ref{relationofxp}) it has a local form. 
From dynamical ansatz (\ref{DynamicalAnsatz})
we find that
\begin{eqnarray}
{\cal Z}(\bm{r},{\cal A}(x_0))={\cal A}(x_0){\cal Z}(\bm{r})\,.
\end{eqnarray}
The inertial vector $\Lambda_a$ has the following form
\begin{align}
	&\Lambda_a=-2i\int d\rho  \left\lbrace \left({\cal Z}^\dagger F_a {\cal Z} \right)\partial_\rho
	\left({\cal Z}^\dagger\partial_\varphi {\cal Z}\right)\right.\notag\\
	&\left.\qquad\qquad\qquad\qquad\qquad
	-\left({\cal Z}^\dagger\partial_\varphi {\cal Z}\right)\partial_\rho\left({\cal Z}^\dagger F_a {\cal Z} \right)\right\rbrace .
\label{InertiaVector}
\end{align}
Explicit form of components of (\ref{InertiaTensor}) and (\ref{InertiaVector}) obtained after imposing  (\ref{ansatz}) is presented in Appendix B.

In virtue of axial symmetry imposed by (\ref{ansatz})  the effective Lagrangian (\ref{effectiveL}) contains the following relevant terms
\begin{align}
&L_{\rm eff}=
\frac{1}{2}\left[I_{11}(\Omega_1^2+\Omega_2^2)+I_{33}\Omega_3^2+2I_{38}\Omega_3\Omega_8+I_{88}\Omega_8^2\right] \notag\\
&\qquad\qquad+\frac{\Theta}{4\pi}\left\{\Lambda_3\Omega_3+\Lambda_8\Omega_8 \right\}-M_{\rm cl}.
\end{align}

\begin{figure}[t]
\includegraphics[width=8cm,clip]{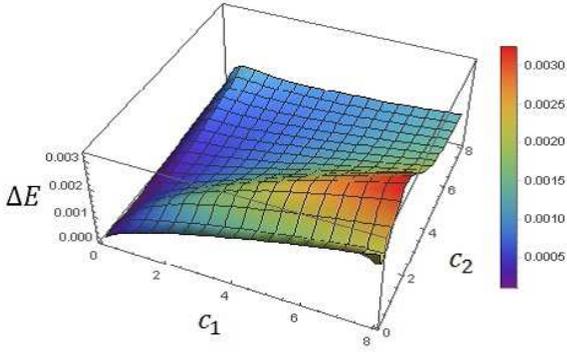}~~~~
\caption{\label{qenergy_holomorphic}The quantum correction of the energy eigenvalue~(\ref{EnergySpectrum}) of the holomorphic 
solutions for the topological charge $Q_{\rm top}=3$, {\it i.e.},$(n_1,n_2)=(2,-1))$ in unit of $(-e^2/4)$. 
The quantum numbers are $(l,k,Y)=(1,1,1)$ and the anyon angle $\Theta=\pi$.}
\end{figure}

The Lagrangian (\ref{effectiveL}) possesses several global continuous symmetries  
that lead to corresponding conserved Noether currents $\mathcal{I}_a,\mathcal{K}_a,\mathcal{J}$, namely 
\begin{eqnarray}
&&(\text{i})~~{\rm left}~SU_L(2): 
~~~~\mathcal{A}\rightarrow e^{-iF_a\xi_a^L}\mathcal{A},\nonumber \\
&&\hspace{1.5cm}\mathcal{I}_a=I_{bc}\Omega_b R_{ac}+\frac{\Theta}{4\pi}\Lambda_a, \\ 
&&(\text{ii})~~{\rm right}~SU_R(2): 
~~~~\mathcal{A}\rightarrow \mathcal{A}e^{iF_a\xi_a^R}, \nonumber \\
&&\hspace{1.5cm}\mathcal{K}_a=-I_{ab}\Omega_b -\frac{\Theta}{4\pi}\Lambda_a,\\
&&(\text{iii})~~{\rm the~circular~symmetry}~X(\rho,\varphi)\rightarrow X(\rho,\varphi+\varphi_0), \nonumber \\
&&~~{\rm or}~~\mathcal{A}\to\mathcal{A}e^{i\bar{F}\varphi_0},~~
\bar{F}:=n_2F_3-\frac{1}{2}(2n_1-n_2)F_4, \\
&&\hspace{1.5cm}\mathcal{J}=n_2\mathcal{K}_3
-\frac{1}{2}(2n_1-n_2)\mathcal{K}_4,
\end{eqnarray}
where the symbol $R_{ab}$ is defined by 
\br
\mathcal{A}\lambda_a\mathcal{A}^\dagger=\lambda_bR_{ba}.
\er
where $\lambda_a$ stands for the Gell-Mann matrices.
Here, $\mathcal{I}_a$ and $\mathcal{K}_a$ are called the coordinate-fixed 
isospin and the body-fixed isospin respectively. $\mathcal{J}$ is a generator of the 
spatial rotation around the third axis. They act on $\mathcal{A}$ as
\begin{equation}
\begin{split}
&[\mathcal{I}_a,\mathcal{A}]=-F_a\mathcal{A},~~
[\mathcal{K}_a,\mathcal{A}]=\mathcal{A}F_a\\
&[\mathcal{J},\mathcal{A}]=\mathcal{A}\bar F\,.
\label{comutationsA}
\end{split} 
\end{equation}
One can construct the explicit form of operators that satisfy (\ref{comutationsA})
\begin{eqnarray}
&&\mathcal{I}_1=i\left(\cos \vartheta_1\cot \vartheta_2\frac{\partial}{\partial \vartheta_1}+\sin \vartheta_1
\frac{\partial}{\partial \vartheta_2}-\frac{\cos \vartheta_1}{\sin \vartheta_2}\frac{\partial}{\partial \vartheta_3}\right), \nonumber \\
&&\mathcal{I}_2=i\left(\sin \vartheta_1\cot \vartheta_2\frac{\partial}{\partial \vartheta_1}-\cos \vartheta_1\frac{\partial}{\partial \vartheta_2}
-\frac{\sin \vartheta_1}{\sin \vartheta_2}\frac{\partial}{\partial \vartheta_3}\right),\nonumber \\
&&\mathcal{I}_3=-i\frac{\partial}{\partial \vartheta_1},\quad \mathcal{I}_8=-i\frac{\partial}{\partial \vartheta_8}, \nonumber \\
&&\mathcal{K}_1=-i\left(\frac{\cos \vartheta_3}{\sin \vartheta_2}\frac{\partial}{\partial \vartheta_1}
-\sin \vartheta_3\frac{\partial}{\partial \vartheta_2}-\cot \vartheta_2\cos \vartheta_3\frac{\partial}{\partial \vartheta_3}\right), \nonumber \\
&&\mathcal{K}_2=i\left(\frac{\sin \vartheta_3}{\sin \vartheta_2}\frac{\partial}{\partial \vartheta_1}
+\cos \vartheta_3\frac{\partial}{\partial \vartheta_2}-\cot \vartheta_2\sin \vartheta_3\frac{\partial}{\partial \vartheta_3} \right), \nonumber \\
&&\mathcal{K}_3=i\frac{\partial}{\partial \vartheta_3},\quad \mathcal{K}_8=i\frac{\partial}{\partial \vartheta_8}, \nonumber \\
&&\mathcal{J}=in_2\frac{\partial}{\partial \vartheta_3}-i\frac{1}{2}(2n_1-n_2)\frac{\partial}{\partial \vartheta_8}.
\end{eqnarray} 
The $SU(2)$ Casimir operator
\br
\bm{\mathcal{I}}^2:=\sum_{i=1}^3\mathcal{I}_i^2=
\sum_{i=1}^3\mathcal{K}_i^2=:\bm{\mathcal{K}}^2,
\er
as well as the generators $\mathcal{I}_3,\mathcal{I}_8,\mathcal{K}_3,\mathcal{K}_8$ 
and $\mathcal{J}$ are diagonalizable. 
The Lagrangian (\ref{effectiveL}) is a function of the four Euler 
angles $\vartheta_i$ and their time derivatives $\dot{\vartheta}_i\equiv\partial_{x_0}\vartheta$,  
{\it i.e.} ${\cal L}_{\rm eff}={\cal L}_{\rm eff}(\vartheta_i,\dot{\vartheta}_i)$. The Legendre  transform of the Lagrangian ${\cal L}_{\rm eff}$ leads to the Hamiltonian ${\cal H}(\vartheta_i,\pi_i):=\pi_i\dot{\vartheta}_i-{\cal L}_{\rm eff}$ which depends on the Euler angles and the canonical momenta
 $\pi_i:=\partial {\cal L}_{\rm eff}/\partial \dot{\vartheta}_i$.
The Hamiltonian takes the form
\begin{align}
&H_{\rm eff}=
M_{\rm cl}+\frac{g_{11}}{2}\left({\cal K}_1^2+{\cal K}_2^2\right) \notag\\
&\qquad+\frac{g_{33}}{2} \biggl({\cal K}_3+\frac{\Theta}{4\pi}\Lambda_3\biggr)^2
+\frac{g_{88}}{2} \biggl({\cal K}_8+\frac{\Theta}{4\pi}\Lambda_8\biggr)^2 \notag\\
&\qquad\qquad\qquad +g_{38} \biggl({\cal K}_3+\frac{\Theta}{4\pi}\Lambda_3\biggr)
\biggl({\cal K}_8+\frac{\Theta}{4\pi}\Lambda_8\biggr)
\end{align}
where we have introduced the components of inverse of inertia tensor $g_{ab}$ whose explicit form in the current case is given by
\begin{eqnarray}
&g_{11}:=\dfrac{1}{I_{11}}~,~~~~~~~~~~~~~~~
g_{33}:=\dfrac{I_{88}}{I_{33}I_{88}-I_{38}^2} ~,\notag\\
&g_{38}:=\dfrac{-I_{38}}{I_{33}I_{88}-I_{38}^2} ~,~~~
g_{88}:=\dfrac{I_{33}}{I_{33}I_{88}-I_{38}^2} \notag~.
\label{momentofinertia}
\end{eqnarray}

The diagonalization problem can be solved using the standard Wigner function
~(for example, \cite{Varshalovich:1988ye})
\br
|lmk;Y\rangle=\mathcal{D}_{m,k}^l(\vartheta_1,\vartheta_2,\vartheta_3)e^{-iY \vartheta_8}|0\rangle
\label{state}
\er
where $l,m,k $ are integer/half-integer and $Y$ has $\frac{1}{3},\frac{2}{3},\cdots$. 
Then the Hamiltonian eigenvalues read 
\begin{align}
&E=
M_{\rm cl}+\frac{g_{11}}{2}\left\{l(l+1)-k^2\right\} \notag\\
&\qquad+\frac{g_{33}}{2} \biggl(k+\frac{\Theta}{4\pi}\Lambda_3\biggr)^2
+\frac{g_{88}}{2} \biggl(Y+\frac{\Theta}{4\pi}\Lambda_8\biggr)^2 \notag\\
&\qquad\qquad\qquad +g_{38} \biggl(k+\frac{\Theta}{4\pi}\Lambda_3\biggr)\biggl(Y+\frac{\Theta}{4\pi}\Lambda_8\biggr)\,.
\label{EnergySpectrum1}
\end{align}
It is convenient to express all the quantum numbers in terms of integer numbers, {\it i.e.}, $l'\equiv 2l, k'\equiv 2k$ and  $Y'\equiv 3Y$. It leads to the following expression for the energy
\begin{align}
&E=
M_{\rm cl}+\frac{\tilde{g}_{11}}{2}\left\{l'(l'+2)-k'^2\right\} \notag\\
&\qquad+\frac{\tilde{g}_{33}}{2} \biggl(k'+\frac{\Theta}{2\pi}\Lambda_3\biggr)^2
+\frac{\tilde{g}_{88}}{2}\biggl(Y'+\frac{3\Theta}{4\pi}\Lambda_8\biggr)^2 \notag\\
&\qquad\qquad\qquad +\tilde{g}_{38} \biggl(k'+\frac{\Theta}{2\pi}\Lambda_3\biggr)\biggl(Y'+\frac{3\Theta}{4\pi}\Lambda_8\biggr)
\label{EnergySpectrum}
\end{align}
where $\tilde{g}_{11}\equiv g_{11}/4,\tilde{g}_{33}\equiv g_{33}/4, \tilde{g}_{38}\equiv g_{38}/6, \tilde{g}_{88}\equiv g_{88}/9$ (in further analysis we shall omit $'$ of $k',l',Y'$ 
for simplicity. )

\begin{figure}[t]
\includegraphics[width=9.0cm,clip]{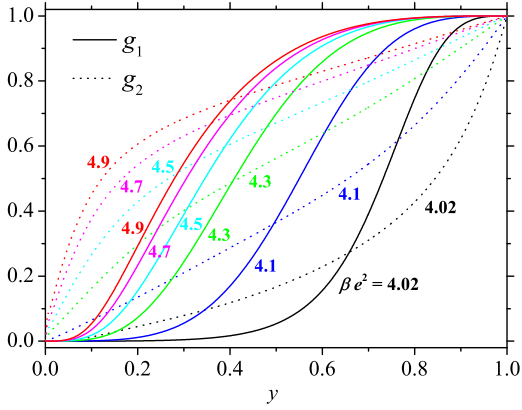}
\caption{\label{csolutions}The classical solutions with 
the topological charge $Q_{\rm top}=3$, {\it i.e.},$(n_1,n_2)=(3,1))$.}
\end{figure}

The derivative expansion of partition function obtained for baby-skyrmions can be straightforwardly applied to the model with $N=2$. In such a case the Dirac operator is of the form
\begin{eqnarray}
i{\cal D}:=i\gamma^\mu (\partial_\mu-iA_\mu)-mX
\end{eqnarray} 
and  real and imaginary parts of the effective  action read
\begin{eqnarray}
&&{\rm Re}S_{\rm eff}=\frac{|m|}{2\pi}\int d^3x{\rm Tr}(\partial_\mu X\partial^{\mu} X)
+O(\partial X^3),
\label{effectiveaction_xr}
\\
&&{\rm Im}S_{\rm eff}=\int d^3x\Bigl(j^\mu (X) A_\mu-\pi\,{\rm sgn}(m){\cal L}_{\rm Hopf}(X)\Bigr)\,.
\label{effectiveaction_xi}
\end{eqnarray}
In analogy to the previous section, one can fix the anyon angle $\Theta$  as $\Theta\equiv \pi\,{\rm sgn}(m)$ 
provided that the vortices are coupled with fermionic field. 
However, since $\Pi_3(\mathbb{C}P^N)$ is trivial, the Hopf term itself ${\cal L}_{\rm Hopf}$ 
is perturbative and the value of the integral depends on the background classical solutions. Consequently, one could not expect that this value became an integer number. 
As a result, the solitons are always anyons even if $\Theta=n\pi, n\in \mathbb{Z}$.

\subsection{Case: $n_1<0$}
In this case the asymptotic value of the Hermitian principal variable $X$ at spatial infinity  reads $X_\infty=\mathrm{diag}(1,1,-1)$. It follows that  four generators of the symmetry $\{F_a\},a=1,2,3,8$ can be chosen as  
\begin{eqnarray}
F_1:=\frac{\lambda_1}{2},~ F_2:=\frac{\lambda_2}{2},~ F_3:=\frac{\lambda_3}{2},~F_8:=\frac{\lambda_8}{\sqrt{3}}\,.
\end{eqnarray}
The quantum Hamiltonian can be diagonalized in the same way as for $n_1>0$ what leads to the energy spectrum (\ref{EnergySpectrum}).

\begin{figure*}[t]
\includegraphics[width=8.5cm,clip]{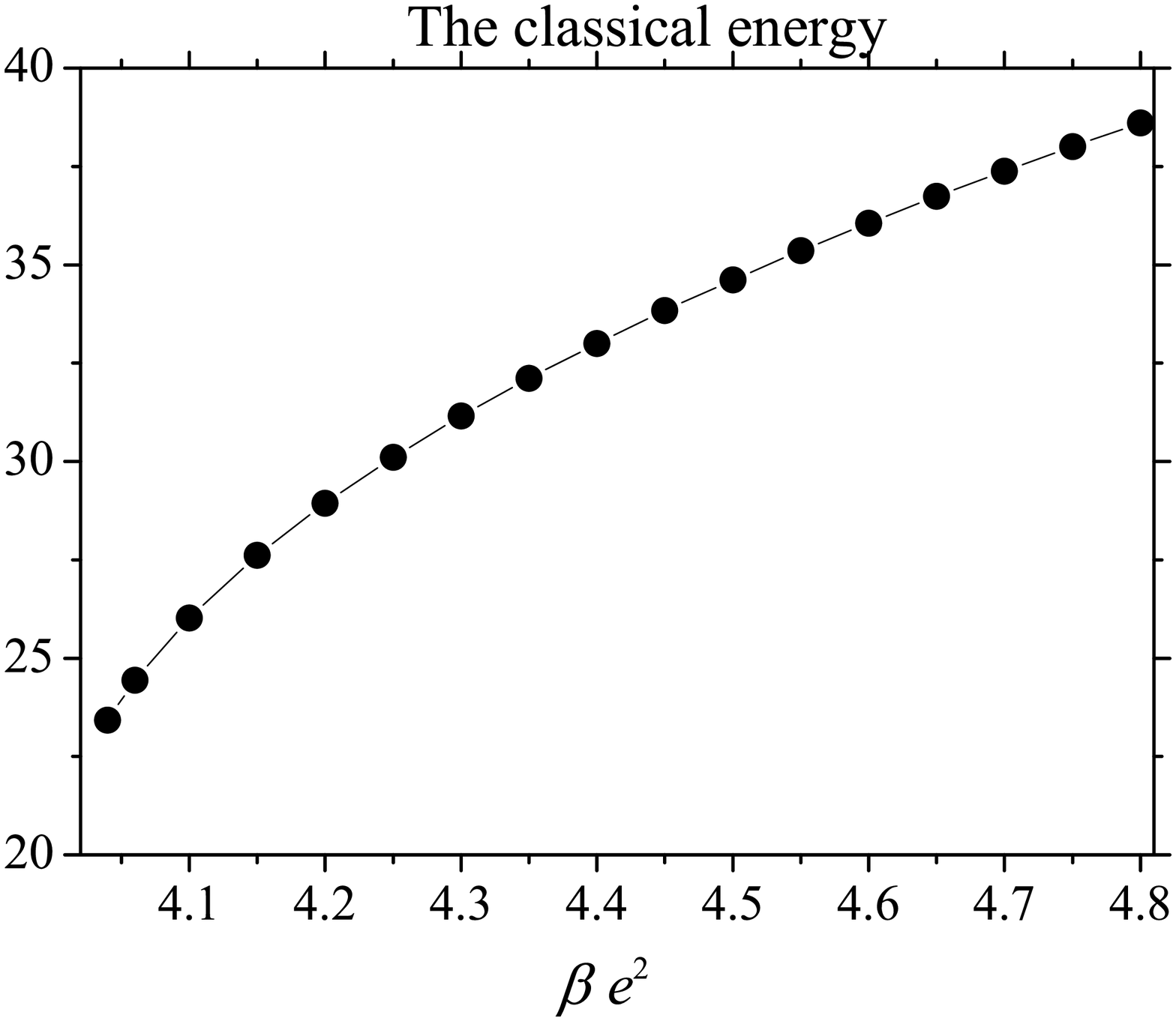}\hspace{-0.5cm}
\includegraphics[width=8.5cm,clip]{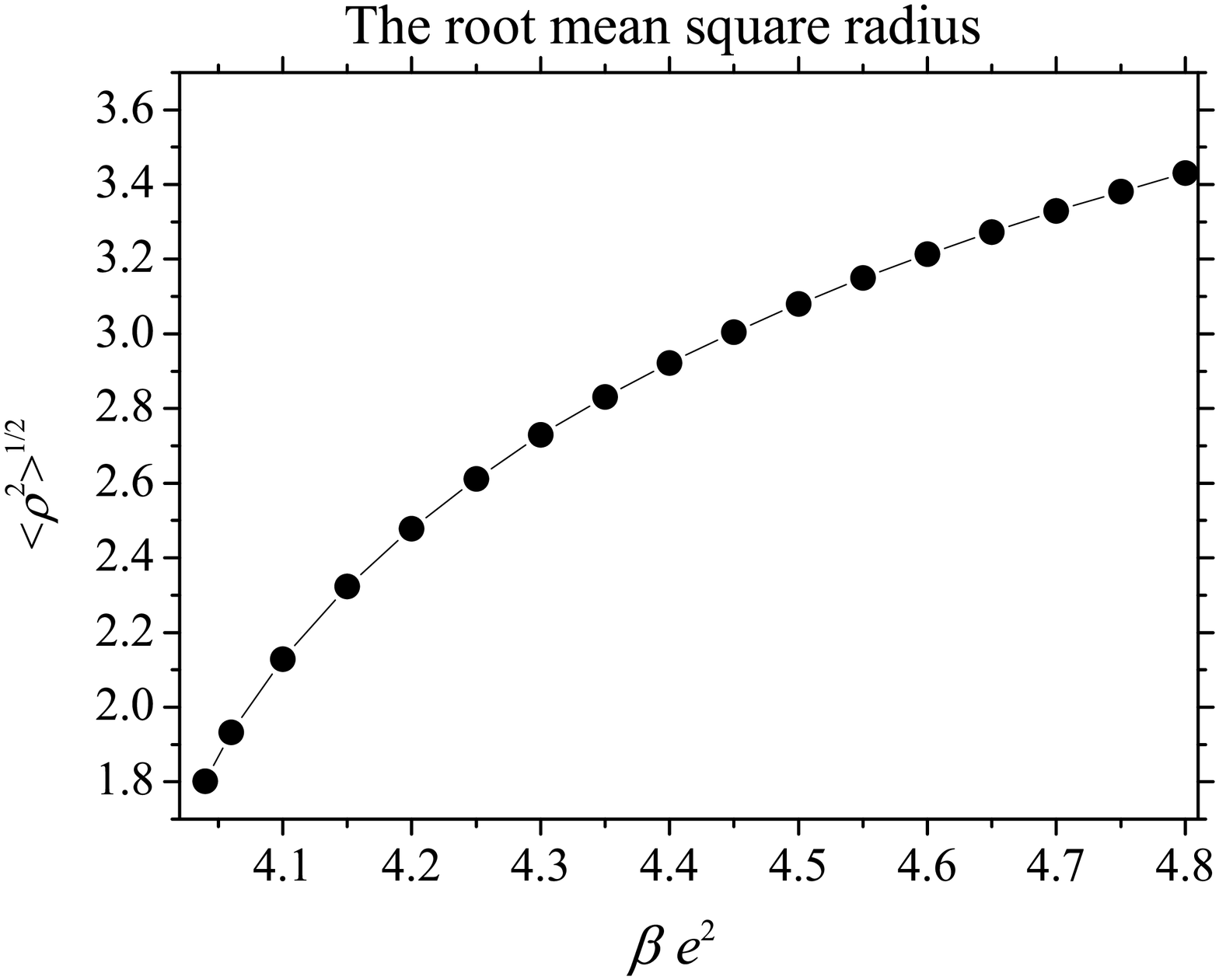}
\caption{\label{energies}The energies (in unit of $4M^2$) and 
the root mean square radius $\sqrt{\langle \rho^2 \rangle }$ (in unit of $-4/M^2e^2$) of the classical solutions 
with the topological charge $Q_{\rm top}=3$, {\it i.e.}, 
 $(n_1,n_2)=(3,1)$.}
\end{figure*}

\section{The analysis}
We shall examine quantum spectra of the model (\ref{actionx})
in the integrable sector $\beta e^2+\gamma e^2=2$ where do exist a class of holomorphic vortex solutions. We choose
\begin{eqnarray}
\beta e^2=4, \gamma e^2=-2.
\end{eqnarray}
in order to satisfy the condition (\ref{properhamiltonian}) simultaneously with the previous one. In the integrable sector the model possesses  exact solutions
\begin{eqnarray}
u_j=c_j\rho^{n_j}e^{in_j\varphi}\label{vortscale}
\label{holomorphic}
\end{eqnarray}
where $c_j$ are arbitrary scale parameters. 
The lowest nontrivial vortex configurations with topological charge (\ref{topologicalcharge}) taking the value
$Q_{\rm top}=2$ are given by  $(n_1,n_2)=(2,1), (1,-1)$. Note that the first term of the inertia tensor (\ref{InertiaTensor})
\begin{eqnarray}
I_{ab}^{Nl\sigma}=\frac{4}{e^2} \int\rho d\rho d\varphi
\mathrm{Tr}\Bigl(\left[F_a,X\right]\left[F_b,X\right]\Bigr) 
\end{eqnarray}
is logarithmically divergent. It means that it has no proper quantum numbers unless one employ a suitable regularization scheme. A similar situation has been identified for  baby-skyrmions in an 
antiferromagnetism~\cite{Vlasii:2012kw}. In such a case the moment of inertia corresponding to the solution with the winding number $n=1$ diverges and because of it no quantized states emerge. The authors have introduced a regularization 
term in order to get the finite value of the integral. 

Here we shall study some configurations of vortices with finite moments of inertia characterized by $Q_{\rm top}=3$. 
The values of the moments of inertia (\ref{InertiaTensor}) for holomorphic solutions are shown in Fig.\ref{momentofinertia_holomorphic}. The moments are shown in dependence on dilatation parameters $(c_1,c_2)$ defined by (\ref{vortscale}). The unit scale is given by $-4/e^2$.
Similarly, the finite components of inertia vector (\ref{InertiaVector}) of the holomorphic solutions have been shown in Fig.\ref{lambda_holomorphic}.
Note that for $c_2\to 0$ the component becomes trivial then the value of 
the vector becomes topological, {\it i.e.}, $\Lambda_3\to -Q_{\rm top}$,  
what is consistent with the analytical calculation for the 
baby-skyrmions (\ref{lambda3_bs}).  
In Fig.\ref{qenergy_holomorphic} 
we plot the dimensionless quantum energy correction $\Delta E$ corresponding 
to the moments of inertia shown in Fig.\ref{momentofinertia_holomorphic}.

\begin{figure*}[t]
\includegraphics[width=19cm,clip]{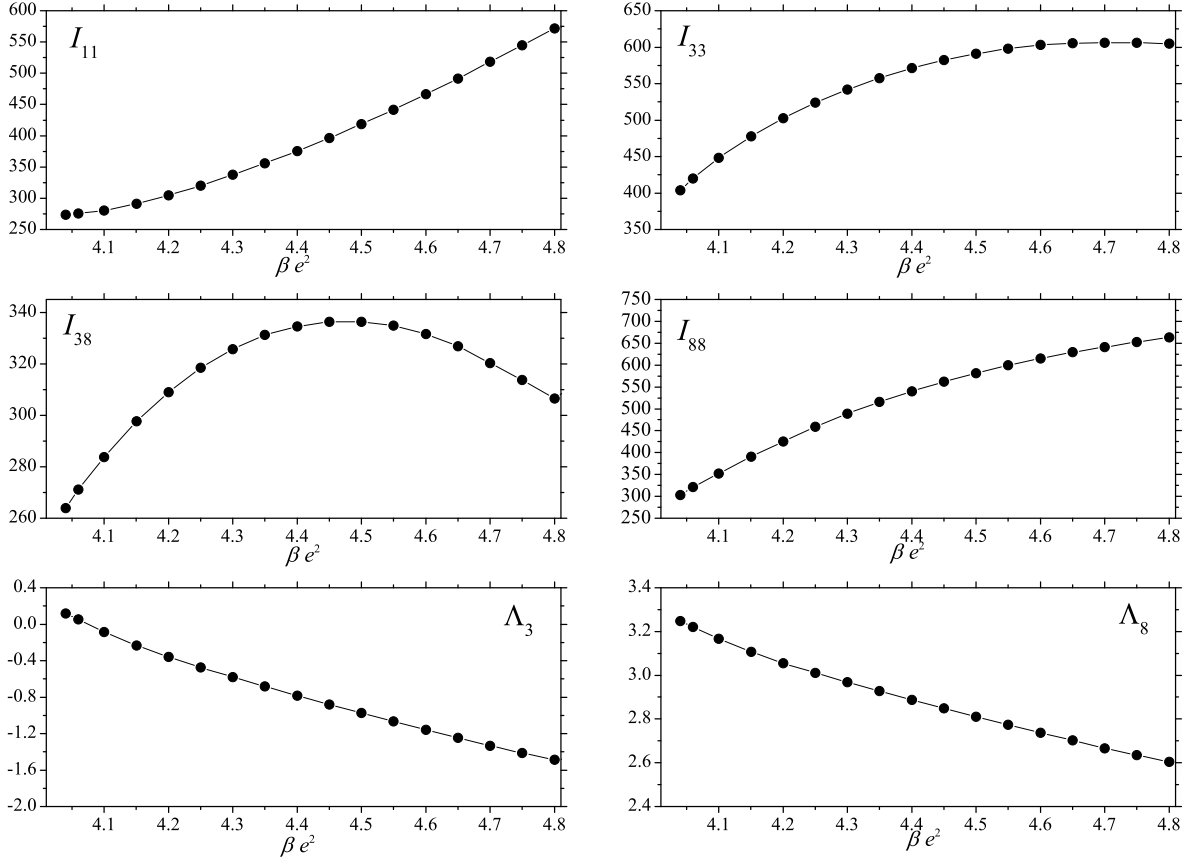}~~
\caption{\label{minertias}Components of the tensor and the vector 
of inertia corresponding to the solutions of Fig.\ref{csolutions}. 
}
\end{figure*}

\begin{figure*}[t]
\includegraphics[width=19cm,clip]{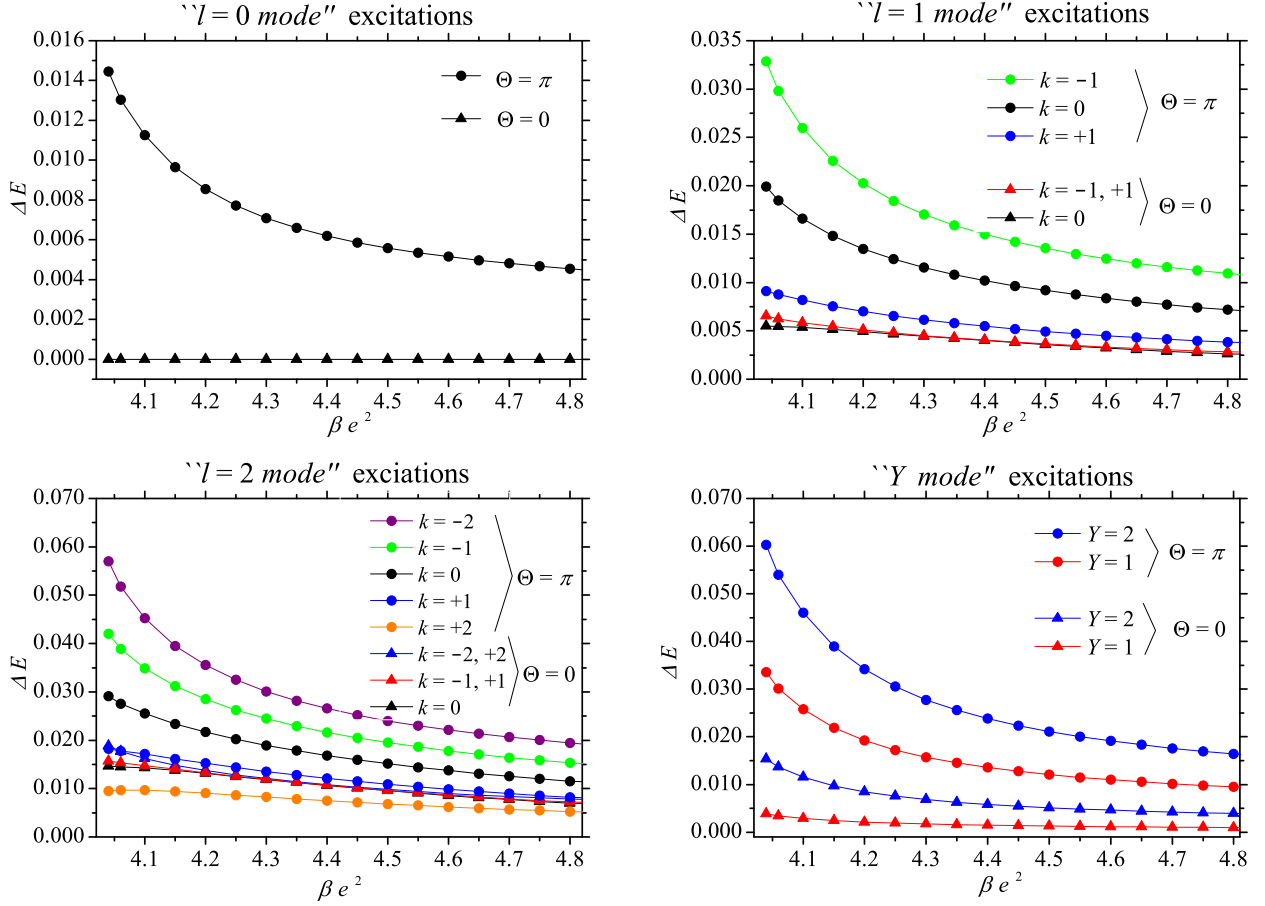}~~
\caption{\label{excitation}The excitation energies
corresponding to the solutions of Fig.\ref{csolutions}. 
We plot the $l=0,1,2$ modes with $Y=0$ and for 
the $Y$ mode we fix $(l,k)=(0,0)$.
}
\end{figure*}

For coupling parameters such that the condition $\beta e^2+\gamma e^2=2$ does not hold the solution is no longer holomorphic.  In such a case the numerical analysis
is required in order to compute the quantum corrections. 
The numerical analysis for the classical solutions of (\ref{equationr}) has been extensively studied in~\cite{Amari:2015sva}. 
Here we shall employ the potential 
\begin{eqnarray}
V=\frac{(1+f_2^2)^2}{(1+f_1^2+f_2^2)^2}\,.
\end{eqnarray}
The corresponding classical solutions for several values of $\beta e^2$ are shown in Fig.\ref{csolutions}. 
In Fig.\ref{energies} we present the corresponding energies and values of the isoscalar root mean square radius
\begin{eqnarray}
\sqrt{\langle \rho^2 \rangle } 
\equiv \biggl[\biggl(-\frac{4}{M^2e^2}\biggr)\int \rho^2j^0(\rho)d\rho\biggr]^{1/2}.
\label{rmsr}
\end{eqnarray} 
The components of the moment of inertia tensors $I_{ab}$ and also 
the inertia vectors $\Lambda_a$ are shown in Fig.\ref{minertias}. 
When $N>1$ there are several excitation modes which are labeled by the 
quantum numbers $l,k,Y$. In Fig.\ref{excitation} we plot the excitation modes 
in dependence on $l,k$ with fixed $Y=0$ 
(``$l$-mode''), and also in dependence on $Y$ with fixed $(l,k)=(0,0)$ (``$Y$-mode''). 

The dimensionless classical energy $\tilde{M}_{\rm cl}\equiv M_{\rm cl}/8\pi M^2$ is 
topological for the holomorphic solutions, {\it i.e.}, it equals to the topological charge $Q_{\rm top}$. 
Clearly, for nonholomorphic solutions the energy deviates from $Q_{\rm top}$, 
however,  it still can be useful to introduce a coupling strength for the quantum correction
$\alpha:=-\frac{e^2}{4}\times \frac{1}{8\pi M^2} = -\frac{e^2}{32\pi M^2}$.
The coupling constants $M^2,e^2$ are some free model parameters, however, their values must be determined by underlying physics.

In order to get some rough idea about properties of quantum excitations, 
it might be instructive to estimate value of the quantum excitations for a given energy scale. For hadronic scale analysis one usually fixes the coefficient of the second order terms $\frac{f_\pi^2}{4}$ in the standard Skyrme model as being equal to the pion decay constant $f_\pi\simeq 64.5 {\rm MeV}$. 
Similarly, we put the coupling constant $M^2$ of the extended Skyrme-Faddeev model as being equal to the pion decay constant $M^2\sim f_\pi\simeq 10^2$ MeV
and we also employ the isoscalar charge density of the nucleon, {\it i.e.},  
$\sqrt{\langle \rho^2\rangle}\simeq 0.7$ fm in (\ref{rmsr}). 
For instance, in the case of our numerical solution with $\beta e^2=4.1$ 
one can easily estimate $-e^2/4\simeq 3.6\times 10^3$ MeV. 
For such a choice of parameters the classical energy is about $M_{\rm cl}\simeq 10.4$ GeV 
and the quantum corrections read 
\begin{eqnarray}
-\frac{e^2}{4}\Delta E &\simeq& 10.4~{\rm MeV}~~``Y~mode''~~(Y=1,\Theta=0), \\
&\simeq& 19.3~{\rm MeV}~~``l~mode''~~(l=1,k=0,\Theta=0).
\end{eqnarray}
The coupling constant takes the value $\alpha\simeq 1.4$. 
Unfortunately, no physical candidate for such small energy excitation for the nucleon are known. One has to stress that the result was obtained as a crude estimation.  Moreover,  the present model is only a two-dimensional mimic of the realistic 3+1 Skyrme model.

Another example of such estimation can be done for an antiferromagnetic material.  A simple estimation of a different type of quantum correction 
in the case of a continuum limit of a Heisenberg model was demonstrated in \cite{rodriquez}. 
In the continuum Heisenberg model, the parameter $M^2$
can be assigned as the exchange coupling constant or a {\it a spin stiffness}.
In antiferromagnetic La$_2$CuO$_4$ the spin wave velocity $c$ 
is of the order $\hbar c\geq 0.04$ eV nm 
and the lattice constant is of the order $a\simeq 0.5$ nm. The exchange coupling constant is roughly $M^2\sim \hbar c/a\simeq 0.1$ eV.   
The soliton size $\sqrt{\langle \rho^2 \rangle}$ is responsible for the size of the excitation then it may be estimates as
$\sqrt{\langle \rho^2 \rangle} \sim a\simeq 0.5$ nm.
In the case of our numerical solution with $\beta e^2=4.1$ we have $-e^2/4\simeq 7.1\times 10^6$ eV.
The classical energy is estimated as $M_{\rm cl}\simeq 10.4$ eV and 
\begin{eqnarray}
-\frac{e^2}{4}\Delta E &\simeq& 20.5~{\rm keV}~~``Y~mode''~~(Y=1,\Theta=0), \\
&\simeq& 38.1~{\rm keV}~~``l~mode''~~(l=1,k=0,\Theta=0).
\end{eqnarray}
The coupling constant $\alpha$ takes the approximate value $\alpha\simeq 2.8\times 10^6$.

Note that presented above estimations were performed in order to get some rough idea about the order of magnitude of excitations.
For more definite analysis one needs further inputs derived from underlying physics.

\section{Generalization to higher $N$}

Though it seems to be certainly involved, a generalization of our scheme to higher $N\geqq 3$ is straightforward. 
In order to simplify the formulation we restrict consideration the case of anyon angle $\Theta=0$. 
We also assume that $n_1$ is the  highest positive integer in the 
sequence $n_i$, $i=1,2,\cdots,N$. 
We present below results for the case $N=3$. 
The generators $\{F^{(3)}_a\}$ read 
\begin{align}
&	F^{(3)}_1=\frac{1}{2}\left(\begin{array}{cc}
0	& 0  \\ 
0	& \lambda_1
\end{array} \right),
F^{(3)}_2=\frac{1}{2}\left(\begin{array}{cc}
0	& 0  \\ 
0	& \lambda_2
\end{array} \right),
	F^{(3)}_3=\frac{1}{2}\left(\begin{array}{cc}
	0	& 0  \\ 
	0	& \lambda_3
	\end{array} \right), \notag \\
&	F^{(3)}_4=\frac{1}{2}\left(\begin{array}{cc}
	0	& 0  \\ 
	0	& \lambda_4
	\end{array} \right),
	F^{(3)}_5=\frac{1}{2}\left(\begin{array}{cc}
	0	& 0  \\ 
	0	& \lambda_5
	\end{array} \right), \notag \\
&	F^{(3)}_6=\frac{1}{2}\left(\begin{array}{cc}
	0	& 0  \\ 
	0	& \lambda_6
	\end{array} \right),
	F^{(3)}_7=\frac{1}{2}\left(\begin{array}{cc}
	0	& 0  \\ 
	0	& \lambda_7
	\end{array} \right), \notag \\
&	F^{(3)}_8=\frac{1}{\sqrt{3}}\left(\begin{array}{cc}
0	& 0  \\ 
0	& \lambda_8
\end{array} \right),
	F^{(3)}_{15}=\frac{1}{4}\left(\begin{array}{cc}
	-3	& 0  \\ 
	0	& I_{3\times 3}
	\end{array} \right).  
	\label{CP3generators}
\end{align}
Following the procedure presented for the $\mathbb{C}P^2$ case, we substitute the dynamical ansatz (\ref{DynamicalAnsatz}) into Lagrangian (\ref{actionx}) and expand the operator $i\mathcal{A}^\dagger \partial_{x_0}{\mathcal{A}}$ in basis of generators $\{F^{(3)}_a\}$ 
\begin{equation}
i\mathcal{A}^\dagger  \partial_{x_0}{\mathcal{A}}=F^{(3)}_a\Omega_a.
\end{equation}
The effective Lagrangian takes the form
\begin{align}
L_{\mathrm{eff}}=
&\sum_{a=1,4,6}\frac{I_{aa}}{2}\left(\Omega_a^2+\Omega_{a+1}^2\right) \notag\\
&\qquad+\sum_{a=3,8,15}\sum_{b=3,8,15}\frac{I_{ab}}{2}\Omega_a\Omega_b  -M_{\mathrm{cl}}. 
\end{align}

The Legendre transformation of the Lagrangian leads to the Hamiltonian 
\begin{align}
H_{\rm eff}=
&\sum_{a=1,4,6}\frac{g_{aa}}{2}\left(\mathcal{K}_a^2+\mathcal{K}_{a+1}^2\right) \notag\\
&\qquad+\sum_{a=3,8,15}\sum_{b=3,8,15}\frac{g_{ab}}{2}\mathcal{K}_a\mathcal{K}_b  +M_{\mathrm{cl}} 
\label{Hamiltonian_CP3}
\end{align}
where $\mathcal{K}_a$ are coordinate-fixed isospin operators and $g_{ab}$ stand for components of inverse of the inertia tensor $I_{ab}$. 
The energy spectrum is obtained after diagonalization of the Hamiltonian (\ref{Hamiltonian_CP3}) in base of  states belonging to a $SU(3)\times U(1)$ irrep and it reads 
\begin{align}
& E_{\mathrm{rot}}=
\sum_{a=1,4,6}\dfrac{g_{aa}}{2}\left\{l_a(l_a+1)-k_a^2\right\} \notag\\
&\qquad\qquad\qquad +\sum_{a=3,8,15} \sum_{b=3,8,15}\dfrac{g_{ab}}{2}k_ak_b, 
\label{Energy_CP3}
\end{align}
where $k_1=k_3, k_4=(2k_3+3k_8)/4$ and $k_6=(2k_3-3k_8)/4$.
The second order Casimir operator of the $SU(3)$ group can be expressed as
\begin{equation}
	C_2(SU(3))=l_1(l_1+1)+l_4(l_4+1)+l_6(l_6+1)-\frac{4k_3^2+3k_8^2}{8}.
\end{equation}

The generalization to an arbitrary $N$ 
is almost straightforward. We can define the $SU(N)\times U(1)$ generators similarly to (\ref{CP3generators}) using a higher $N$ generalization of Gell-Mann matrices $\lambda^{N}_a$ which are the standard $SU(N)$ generators.
We take the diagonal components of $SU(N)$ part (counterparts of $F_3$ and $F_8$)  
\begin{eqnarray}
&&	F^{(N)}_{a^2-1}=\dfrac{1}{a}\left(\begin{array}{cc}
0	& 0 \\ 
0	& \lambda^{N}_{a^2-1}
	\end{array} \right) \qquad a=2,3,\cdots,N ~,  
\end{eqnarray}
and the off diagonal ones, like $F_1$ and $F_4$, in the form
\begin{eqnarray}
&&		F^{(N)}_b=\dfrac{1}{2}\left(\begin{array}{cc}
	0	& 0 \\ 
	0	& \lambda^{N}_b
	\end{array} \right), \notag
	\end{eqnarray}
 where $b$ are integer numbers from $1$ to $N^2-1$ excluding $a^2-1$.
 The last $U(1)$ generator is defined as 
	\begin{eqnarray}
	&&	F^{(N)}_{N^2-2N}=\dfrac{1}{N+1}\left(\begin{array}{cc}
	-N	& 0 \\ 
	0	& I_{N\times N}
	\end{array} \right) .
\end{eqnarray}
After lengthy calculation involving generators one gets the quantum energy formula 
\begin{align}
	& E_{\mathrm{rot}}=
	\sum_{i=1}^N\sum_{j=1}^{i-1}\frac{g_{\xi\xi}}{2}\left\{l_\xi(l_\xi+1)-k_\xi^2\right\}  \notag\\
	&\qquad\qquad\qquad+\sum_{i=1}^N\sum_{j=1}^N \dfrac{g_{\eta_i\eta_j}}{2}k_{\eta_i}k_{\eta_j}
	\label{CPN energy}
\end{align} 
where $\xi=(i-1)^2+2(j-1)$ and $\eta_i=i(i+2)$. Symbols $l_\xi$ and $k_{\eta_i}$ represent independent quantum numbers whereas $k_\xi$ are given in terms of $k_{\eta_i}$ according to 
\begin{equation}
	k_\xi=\frac{1}{2}\left(-k_{\eta_{(j-1)}} +\sum_{h=j}^{i} \frac{1}{h}k_{\eta_h}+k_{\eta_i}   \right)
\end{equation}
with $k_0=0$.
Note that $\xi$ and $\eta_i$ describe the numerical sequence
\begin{align}
&~\{\xi\}~:~1,~4,~6,~9,~11,~13,~16,~18,~20,~22,~\cdots \\
&\{\eta_i\}~:~3,~8,~15,~24~\cdots~.
\end{align}
The energy formula \eqref{CPN energy} contains the first $N(N-1)/2$ terms of $\{\xi\}$ and the $N$ terms of $\{\eta_i\}$, {\it e.g.}
\begin{align}
&\text{for}~N=2~:\quad\xi=1, ~~~~~~~~~~~~\eta_i=3,8 \notag\\
&\text{for}~N=3~:\quad\xi=1,4,6 ~~~~~~\eta_i=3,8,15. \notag
\end{align}
One can check that for this values expression \eqref{CPN energy} certainly reduces to \eqref{EnergySpectrum1} for $N=2$ (and $\Theta=0$) 
and to \eqref{Energy_CP3} for $N=3$.

\section{Summary}

The present paper aims at the problem of quantum spectra of solutions in the extended $\mathbb{C}P^N$ Skyrme-Faddeev model. 
In order to obtain the quantum energy spectra of excitations we applied the method of collective coordinate quantization 
based on a rigid body approximation. Further, within this approximation, we discussed 
spin statistics of the $\mathbb{C}P^N$ soliton taking into account the Hopf Lagrangian.
 According to discussion presented in previous papers \cite{Wilczek:1983cy,Wu:1984kd,Bar:2003ip,Jaroszewicz:1985ip}, 
for $N=1$ the Hopf Lagrangian is topological so the solitons are quantized as anyons
with the angle $\Theta$. On the other hand, for $N>1$, the Hopf term is perturbative thus the solutions became always anyons. 

A fermionic effective model coupled with 
the skyrmion of a constant gap $mX$ has been examined \cite{Abanov:2000ea,Abanov:2001iz}. 
After integrating out the Dirac 
field the resulting effective Lagrangian contains the 
Skyrme-Faddeev model plus some topological terms. 
For $N=1$, the anyon angle $\Theta$ became fixed, then in contrary to the previous
papers, the solitons cannot be anyon at all, whereas for $N>1$, the $\Theta$  became  
fixed again but since in this case the Hopf term is perturbative then the solution became an anyon. 

The further part contains the study of excitation energy of the solutions. 
The excited energy for $N=1$ is described by (a third component of) angular momentum $S:=\ell-\frac{\Theta}{2\pi}$. 
According to \cite{Kovner:1989wd}, for $\Theta=\pi$ the baby skyrmions 
are fermions and the ground state is twice degenerated. For $N>1$ the situation is quite different. 
The solutions are always anyon type because of the fact that the Hopf term is no longer topological. It follows that there are
no degeneracy. The excitations are parametrized by 
three numbers $l,k,Y$,  so $S_3:=k+\frac{\Theta\Lambda_3}{2\pi}, S_8:=Y+\frac{3\Theta\Lambda_8}{4\pi}$ 
are components of the anyonic angular momenta.
The paper contain plots of some energy levels in dependence on the model
parameters $\beta e^2$. The presented values of the energy are dimensionless.

We gave some rough estimations of typical energy scales being of order of few dozens 
MeV for a hadronic scale and of order of few dozens keV for a condensed matter scale. 
The energy excitation compared with the classical solution energy is subtle for a 
hadronic scale whereas it is virtually too big in the case of the condensed matter example.   
Such discrepancy can be understand to some extent. 
For instance, if we choose a solution parameters $Q_{\rm  top}=3$ and $\beta e^2=4.1$   
then estimation of the classical energy gives
$M_{\rm cl}\simeq 1.1\times 10^2 M^2$ and for the quantum excitation energy of the lowest $Y$ mode it has the value $-e^2/4\times \Delta 
E\simeq 5.5\times 10^2/(M^2 \langle \rho^2\rangle )$. The ratio of this two energies is given by
\begin{eqnarray}
\frac{-\frac{e^2}{4}\Delta E}{M_{\rm cl}}\simeq 5\times\frac{1}{M^4\langle \rho^2\rangle}
\end{eqnarray}
where dimensions of $M^2, \sqrt{\langle \rho^2\rangle}$ are [eV,~nm] or [MeV,~fm].
One can easily see that a huge discrepancy of the classical/quantum energy 
for the antiferromagnetic material is fixed almost by a value of the spin wave velocity. 
However, some systems may support different values of  
the parameters $M^2$ and $\sqrt{\langle \rho^2\rangle}$. 
The systems with higher values of coupling constant $M^2$ (which determines the energy scale) 
and with a large characteristic excitation size $ \sqrt{\langle \rho^2\rangle}$ 
can support the existence of excitations whose energy is comparable with 
the classical energy. We shall leave this problem for the future.

One has to bear in mind that the collective coordinate quantization is an approximated method. 
An alternative method which  can be applied to quantization of the vortex systems is a
 canonical quantization method. Such approach would be more suitable in full understanding of 
the quantum aspects of the model. The work is in progress and its results will be reported in a subsequent paper \cite{amari_un}.

\vskip 0.5cm\noindent
{\bf Acknowledgement}

The authors would like to thank L. A. Ferreira for discussions and comments. 
We thank to A. Nakamula and K. Toda for their useful comments. 
Also we are grateful to Y. Fukumoto for giving us many valuable information. 

\appendix
\section{The derivative expansion of the fermionic model}
The method of perturbation for the partition function~(\ref{partition function})
is quite common and widely examined for the analysis of the spin-statistics of 
solitons coupled with fermions. 
Here we employ notation used in \cite{Diakonov:1987ty} with the $\mathbb{C}P^1$ principal variable \eqref{CP1 principal}.
The partition function is given by the integral
\begin{eqnarray}
\Gamma=\int {\cal D}\psi{\cal D}\bar{\psi} e^{\int d^3x \bar{\psi}iD\psi}
=\det iD
\end{eqnarray}
where $\psi$ and $\bar\psi$ are Dirac fields, $A_\mu$ is a U(1) gauge field and $iD=i\gdhi+\gdAi-mX$. 
The gamma matrices are defined as $\gamma^\mu=-i\sigma^\mu$. 
The effective action $S_{\rm eff}=\ln \det iD$ can be split in its real and imaginary part
\begin{eqnarray}
{\rm Re}S_{\rm eff}=\frac{1}{2}\ln \det D^\dagger D, ~~\qquad
{\rm Im}S_{\rm eff}=\frac{1}{2i}\ln \det\frac{iD}{-iD^\dagger}\,.
\end{eqnarray}
For $A_\mu\to 0$ one can easily see that
\begin{eqnarray}
&&D^\dagger D=-\partial^2+m^2+im\gdhi X, \nonumber \\
&&DD^\dagger =-\partial^2+m^2-im\gdhi X\,.
\end{eqnarray}
For the variation
$D\to D+\delta D, D^\dagger\to D^\dagger+\delta D^\dagger$, the real part 
of the effective action in $A_\mu\to 0$ is
\pagebreak
\begin{eqnarray}
&&\delta{\rm Re} S_{\rm eff}= 
\frac{1}{2}{\rm Sp}
\biggl(\frac{1}{D^\dagger D}D^\dagger \delta D+\frac{1}{DD^\dagger}D \delta D^\dagger\biggr) 
\nonumber \\
&&=\frac{1}{2}
\int d^3x\int \frac{d^3k}{(2\pi)^3}e^{-ik\cdot x} \nonumber \\
&&\times {\rm Tr}\biggl[
(-\partial^2+m^2+im\gdhi X)^{-1}(im\gdhi \delta X+m^2X\delta X) \nonumber \\
&&+(-\partial^2+m^2-im\gdhi X)^{-1}(-im\gdhi \delta X+m^2X\delta X)
\biggr]e^{ik\cdot x} \nonumber \\
&&=\frac{1}{2}\int d^3x\int \frac{d^3k}{(2\pi)^3}\nonumber \\
&&\times {\rm Tr}\biggl[
(k^2+m^2-2ik\partial-\partial^2+im\gdhi X)^{-1}(im\gdhi \delta X+m^2X\delta X) \nonumber \\
&&+(k^2+m^2-2ik\partial-\partial^2-im\gdhi X)^{-1}(-im\gdhi \delta X+m^2X\delta X) \nonumber \\
\end{eqnarray}
where ${\rm Sp}$ stands for a full trace containing a functional and also 
a matrix trace involving the flavor and the spinor indices and {\rm Tr} stands for usual matrix trace. 
Expanding the above expression in powers of $2ik\partial+\partial^2$ and $\gdhi X$ one gets the lowest nonzero term
\begin{eqnarray}
\delta S^{(2)}_{\rm Re}=\frac{|m|}{8\pi}\int d^3x {\rm Tr}(\gdhi X\gdhi\delta X)\,.
\end{eqnarray}
After taking the spinor trace and switching to the Minkowski metric one gets
the action (\ref{effectiveaction_fr}). 

For the variation of the imaginary part
\begin{eqnarray}
\delta{\rm Im} S_{\rm eff}=\frac{1}{2i}
{\rm Sp}\biggl(\frac{1}{D^\dagger D}D^\dagger \delta D-\frac{1}{DD^\dagger}D \delta D^\dagger\biggr) 
\end{eqnarray}
the calculation is almost similar and the first nonzero component contains product of the three derivatives
\begin{eqnarray}
\delta{\rm Im} S_{\rm eff}=-\frac{{\rm sgn}(m)}{32\pi}\int d^3x \epsilon^{\mu\nu\lambda}
{\rm Tr}(\partial_\mu X\partial_\nu X\partial_\lambda X X\delta X)\,.
\end{eqnarray}
In terms of new variable $a_{\mu}:=-iZ^\dagger \partial_\mu Z$ the last formula can be written as
\begin{eqnarray}
\delta {\rm Im}S^{(3)}_{\rm eff}=\frac{{\rm sgn}(m)}{2\pi}\int d^3x \epsilon^{\mu\nu\lambda}\delta a_\mu \partial_\nu a_\lambda\,.
\end{eqnarray}
As it was argued in \cite{Abanov:2000ea}, the term itself should be zero, 
{\it i.e.}, ${\rm Im}S^{(3)}_{\rm eff}=0$  in the pertubative calculation because the 
homotopy group $\pi_3(S^2)=\mathbb{Z}$ is nontrivial. 
In order to find the form for $N=1$, we generalize the model into $N\geqq 2$ such as $ Z=(z_1,z_2,0,\cdots,0)^T$. In the case when
the homotopy $\pi_3(CP^N)$ is trivial one gets 
\begin{eqnarray}
{\rm Im}S^{(3)}_{\rm eff}=\frac{{\rm sgn}(m)}{4\pi}\int d^3x \epsilon^{\mu\nu\lambda} a_\mu \partial_\nu a_\lambda\,.
\label{Hopflag}
\end{eqnarray}
For $N>1$, the above manipulation is not a trick and we directly obtain the form (\ref{Hopflag})
with $a_\mu:=-i{\cal Z}^\dagger\partial_\mu {\cal Z}$.  

For $A_\mu\neq 0$ we also have the two derivative component
\begin{eqnarray}
\delta {\rm Im}S^{(2)}_{\rm eff}|_{A_\mu\neq 0}
=-\frac{1}{16\pi i}\delta A_\mu \int d^3x\epsilon^{\mu\nu\lambda}
{\rm Tr}(\partial_\nu X\partial_\lambda X X)\,.
\end{eqnarray}
Both contribute to the final expression (\ref{effectiveaction_fi}). 

\begin{widetext}
\section{Inertia tensor $I_{ab}$ and inertia vector $\Lambda_a$ for $\mathbb{C}P^2$}
In this appendix we give the explicit form of the relevant components of inertia tensor $I_{ab}$ \eqref{InertiaTensor} 
and inertia vector $\Lambda_a$ \eqref{InertiaVector} written explicitly by the radial profile functions $f_1(\rho),f_2(\rho)$
and their derivatives $f'_1:=\frac{df_1}{d\rho},f'_2:=\frac{df_2}{d\rho}$. 
They read
	\begin{align}
	I_{11}&=\frac{8\pi}{e^2}\int \rho d\rho \left[
	-\frac{  \left(1+f_1{}^2 +f_1{}^2f_2{}^2+f_2{}^4\right)}
	{\left(1+f_1{}^2+f_2{}^2\right){}^2} \right.\notag\\
	&+\frac{4  \left\{1+f_1{}^2 \left( 1+8f_2{}^2+f_2{}^4\right)+f_2{}^2+f_2{}^4+f_2{}^6\right\} f_1'{}^2}
	{\left(1+f_1{}^2+f_2{}^2\right){}^4}  
	-\frac{8  f_1 f_2 \left\{4+f_1{}^2 \left(f_2{}^2+4\right)-3f_2{}^2+f_2{}^4\right\} f_1' f_2'}
	{\left(1+f_1{}^2+f_2{}^2\right){}^4}     \notag\\
	&+\frac{4  \left\{\left(f_2{}^2+2\right) f_1{}^4+\left(f_2{}^4-3 f_2{}^2+4\right)
		f_1{}^2+2 \left(f_2{}^2-1\right){}^2\right\} f_2'{}^2}
	{\left(1+f_1{}^2+f_2{}^2\right){}^4}
	+\frac{4  \left\{f_2{}^4 \left(\left(n_1-n_2\right){}^2 f_1{}^2+2 n_2^2\right)+2 n_2^2 \left(1+f_1{}^2\right) f_2{}^2+n_1^2 f_1{}^2\right\}}
	{r^2 \left(1+f_1{}^2+f_2{}^2\right){}^3}
	\notag\\
	&+2\beta  e^2\left\{
	-\frac{2   \left(1+f_1{}^2 +f_1{}^2f_2{}^2+f_2{}^4\right)\left\{\left(1+f_2{}^2\right)f_1'{}^2-2  f_1 f_2  f_1' f_2'+\left(1+f_1{}^2\right)f_2'{}^2\right\} }{\left(1+f_1{}^2+f_2{}^2\right){}^4}
	\right.\notag\\
	&~~~~-\frac{2  \left(1+f_1{}^2 +f_1{}^2f_2{}^2+f_2{}^4\right) \left\{n_1^2 f_1{}^2+ \left(n_1-n_2\right){}^2 f_1{}^2f_2{}^2 +n_2^2f_2{}^2 \right\}}{r^2 \left(1+f_1{}^2+f_2{}^2\right){}^4}
	\left.\left.
	+\frac{  f_2{}^2 \left\{2 n_1 f_1{}^2-n_2 \left(1+f_1{}^2-f_2{}^2\right)\right\}{}^2}
	{r^2 \left(1+f_1{}^2+f_2{}^2\right){}^4}
	\right\}
	\right]\,,\\
I_{33}&=\frac{8\pi}{e^2}\int \rho d\rho \left[-\frac{  \left\{\left(1+f_2{}^2\right) f_1{}^2+4 f_2{}^2\right\}}
{\left(1+f_1{}^2+f_2{}^2\right){}^2}    \right.
+\frac{16   \left\{-f_1{}^2 \left(1-f_2{}^2-f_2{}^4\right)+f_2{}^2+f_2{}^4\right\} f_1'{}^2}{\left(1+f_1{}^2+f_2{}^2\right){}^4}
\notag \\
&+\frac{16   f_1 f_2 \left\{f_1{}^2-\left(2 f_1{}^2+5\right)f_2{}^2+3\right\} f_1' f_2'}
{\left(1+f_1{}^2+f_2{}^2\right){}^4}
+\frac{4  \left\{f_1{}^2+f_1{}^4+\left(16+17 f_1{}^2+4 f_1{}^4\right) f_2{}^2\right\}
	f_2'{}^2}
{\left(1+f_1{}^2+f_2{}^2\right){}^4} 
+\frac{4  \left(2 n_1-n_2\right){}^2 f_1{}^2 f_2{}^2}
{r^2 \left(1+f_1{}^2+f_2{}^2\right){}^3}   
\notag\\
&+4\beta e^2
\left\{-\frac{  \left\{\left(1+f_2{}^2\right) f_1{}^2+4 f_2{}^2\right\}\left\{\left(1+f_2{}^2\right)f_1'{}^2-2  f_1 f_2  f_1' f_2'+\left(1+f_1{}^2\right)f_2'{}^2\right\} }
{\left(1+f_1{}^2+f_2{}^2\right){}^4}       
\right. \notag\\
&-\frac{ \left\{\left(1+f_2{}^2\right) f_1{}^2+4 f_2{}^2\right\} \left\{n_1^2 \left(1+f_2{}^2\right) f_1{}^2-2 n_1 n_2 f_1{}^2 f_2{}^2+n_2^2 \left(1+f_1{}^2\right) f_2{}^2\right\}}
{r^2\left(1+f_1{}^2+f_2{}^2\right){}^4} 
\left.\left.
+\frac{ \left\{n_1 f_1{}^2 \left(1-f_2{}^2\right)+n_2 \left(2+f_1{}^2\right) f_2{}^2\right\}{}^2}{r^2 \left(1+f_1{}^2+f_2^2\right)^4}  \right\}\right]\,,	
\\
I_{38}&=\frac{16\pi}{e^2}\int \rho d\rho \left[\frac{  f_1{}^2 \left(1-f_2{}^2\right)}{\left(1+f_1{}^2+f_2{}^2\right){}^2}
\right.
-\frac{16 f_1{}^2 \left(1-f_2{}^4\right) f_1'{}^2}{\left(1+f_1{}^2+f_2{}^2\right){}^4}
+\frac{8  f_1 f_2 \left\{f_1{}^2-\left(4 f_1{}^2+3\right) f_2{}^2-3\right\} f_1'
	f_2'}{\left(1+f_1{}^2+f_2{}^2\right){}^4}
\notag\\
&-\frac{4  f_1{}^2 \left\{1+\left(1-4 f_2{}^2\right) f_1{}^2-7 f_2{}^2\right\} f_2'{}^2}{\left(1+f_1{}^2+f_2{}^2\right){}^4}
+\frac{4  \left(2 n_1-n_2\right) n_2 f_1{}^2 f_2{}^2}{r^2 \left(1+f_1{}^2+f_2{}^2\right){}^3}
\notag\\ 
&+4\beta  e^2\left\{
\frac{  \left(1-f_2{}^2\right)  f_1{}^2
	\left\{\left(1+f_2{}^2\right)f_1'{}^2-2  f_1 f_2  f_1' f_2'+\left(1+f_1{}^2\right)f_2'{}^2\right\}}
{\left(1+f_1{}^2+f_2{}^2\right){}^4}
\right.\notag\\
&~~~~+\frac{  f_1{}^2 \left(1-f_2{}^2\right) \left\{n_1^2 \left(1+f_2{}^2\right) f_1{}^2-2 n_1 n_2 f_1{}^2 f_2{}^2+n_2^2 \left(1+f_1{}^2\right) f_2{}^2\right\}}{r^2
	\left(1+f_1{}^2+f_2{}^2\right){}^4}
\notag\\
&\left.\left.
~~~~-\frac{  f_1{}^2 \left(n_1 f_1{}^2 \left(1-f_2{}^2\right)+n_2 \left(f_1{}^2+2\right) f_2{}^2\right) \left(n_1 \left(1+f_2{}^2\right)-n_2 f_2{}^2\right)}{r^2 \left(1+f_1{}^2+f_2{}^2\right){}^4}
\right\}
\right]\,,
\\
I_{88}&=\frac{32\pi}{e^2}\int \rho d\rho \left[-\frac{  f_1{}^2 \left(1+f_2{}^2\right)}{\left(1+f_1{}^2+f_2{}^2\right){}^2}
\right.
-\frac{32   \left(1+f_2{}^2\right) f_1{}^3f_2 f_1' f_2'}{\left(1+f_1{}^2+f_2{}^2\right){}^4}
+\frac{16  \left(1+f_2{}^2\right){}^2 f_1{}^2
	f_1'{}^2}{\left(1+f_1{}^2+f_2{}^2\right){}^4}\notag\\ 
&+\frac{4  \left( 1+f_1{}^2 1+f_2{}^2+4f_1{}^2f_2{}^2\right) f_1{}^2 f_2'{}^2}{\left(1+f_1{}^2+f_2{}^2\right){}^4}
+\frac{4  n_2^2 f_1{}^2 f_2{}^2}{r^2 \left(1+f_1{}^2+f_2{}^2\right){}^3}\notag\\
&+4\beta  e^2\left\{
-\frac{  f_1{}^2\left(1+f_2{}^2\right)\left\{\left(1+f_2{}^2\right)f_1'{}^2-2  f_1 f_2  f_1' f_2'+\left(1+f_1{}^2\right)f_2'{}^2\right\}}{\left(1+f_1{}^2+f_2{}^2\right){}^4}
\right.\notag\\
&~~~~-\frac{  f_1{}^2 \left(1+f_2{}^2\right) \left\{n_1^2 \left(1+f_2{}^2\right) f_1{}^2-2 n_1 n_2 f_1{}^2 f_2{}^2+n_2^2 \left(f_1{}^2+1\right) f_2{}^2\right\}}
{r^2\left(1+f_1{}^2+f_2{}^2\right){}^4}
\left.\left.
+\frac{  f_1{}^4 \left\{n_1 \left(1+f_2{}^2\right)-n_2 f_2{}^2\right\}{}^2}{r^2 \left(1+f_1{}^2+f_2{}^2\right){}^4}
\right\}
\right]\,,
\end{align}
and
\begin{align}
&\Lambda_3=-2\int d\rho\frac{n_1 f_1f_1'+n_2 f_2 f_2'+n_1 f_1f_2 \left(f_1 f_2'-f_2 f_1'\right)}
{\left(1+f_1{}^2+f_2{}^2\right){}^2}\,,	
\\
&\Lambda_8=4\int d\rho\frac{ n_1  \left(1+f_2{}^2\right) f_1f_1'+n_2 \left(1-2f_1{}^2\right)
f_2f_2'-f_1f_2\left( n_1f_1f_2'-2n_2 f_2f_1'\right)}{3 \left(1+f_1{}^2+f_2{}^2\right){}^2}\,.
\end{align}
\end{widetext}


\begin{thebibliography}{0}
\bibitem{sf}
  L.~D.~Faddeev, 
  Princeton preprint IAS Print-75-QS70,
  1975;\\
  {\it in 40 Years in Mathematical Physics}, (World
  Scientific, 1995); \\
 L.~D.~Faddeev and A.~J.~Niemi,
  Knots and particles,
  Nature {\bf 387}, 58 (1997)
  [arXiv:hep-th/9610193];
\\
 P.~Sutcliffe,
  Knots in the Skyrme-Faddeev model,
  Proc.\ Roy.\ Soc.\ Lond.\  A {\bf 463}, 3001 (2007)
  [arXiv:0705.1468 [hep-th]];
\\
  J.~Hietarinta and P.~Salo,
  Faddeev-Hopf knots: Dynamics of linked un-knots,
  Phys.\ Lett.\  B {\bf 451}, 60 (1999)
  [arXiv:hep-th/9811053];
\\
  J.~Hietarinta and P.~Salo,
  Ground state in the Faddeev-Skyrme model,
  Phys.\ Rev.\  D {\bf 62}, 081701 (2000).


\bibitem{coleman}
S.~R.~Coleman,
  Quantum sine-Gordon equation as the massive Thirring model,
  Phys.\ Rev.\  D {\bf 11}, 2088 (1975);\\
S.~Mandelstam,
  Soliton operators for the quantized sine-Gordon equation,
  Phys.\ Rev.\  D {\bf 11}, 3026 (1975).

\bibitem{vortexlaf} 
  L.~A.~Ferreira,
  Exact vortex solutions in an extended Skyrme-Faddeev model,
JHEP {\bf 05}, 001 (2009), 
  arXiv:0809.4303 [hep-th].
  
 \bibitem{gies} 
  H.~Gies, 
 Wilsonian effective action for SU(2) Yang-Mills theory with
Phys.\ Rev.\ D {\bf 63}, 125023 (2001), hep-th/0102026


\bibitem{fk} 
L. A. Ferreira, P. Klimas, 
 Exact vortex  solutions in a $CP^N$ Skyrme-Faddeev type model, 
JHEP {\bf 10}, 008 (2010) [arXiv: 1007.1667]
  
\bibitem{fkz} 
L. A. Ferreira, P. Klimas, and W. J. Zakrzewski,
 Some properties of (3+1) dimensional vortex solutions in the extended $CP^N$ Skyrme-Faddeev model, 
JHEP {\bf 12}, 098 (2011).

\bibitem{Amari:2015sva} 
  Y.~Amari, P.~Klimas, N.~Sawado and Y.~Tamaki,
  Potentials and the vortex solutions in the $CP^N$ Skyrme-Faddeev model,
  Phys.\ Rev.\ D {\bf 92}, no. 4, 045007 (2015)
  [arXiv:1504.02848 [hep-th]].



\bibitem{Haldane:1983ru} 
  F.~D.~M.~Haldane,
  Nonlinear field theory of large spin Heisenberg antiferromagnets. Semiclassically quantized solitons of the one-dimensional easy Axis Neel state,
  Phys.\ Rev.\ Lett.\  {\bf 50}, 1153 (1983).

\bibitem{Bowick:1985ua} 
  M.~J.~Bowick, D.~Karabali and L.~C.~R.~Wijewardhana,
  Fractional Spin via Canonical Quantization of the O(3) Nonlinear Sigma Model,
  Nucl.\ Phys.\ B {\bf 271}, 417 (1986).

\bibitem{Kovner:1989wd} 
  A.~Kovner,
  Canonical Quantization of the {CP}$^{N}$ Model With a $\Theta$ Term,
  Phys.\ Lett.\ B {\bf 224}, 299 (1989).

\bibitem{rodriquez}
  J.~P.~Rodriguez, 
  Quantized topological point defects in two-dimensional antiferromagnets, 
  Phys.\ Rev.\ B {\bf 39}, 2906 (1989); {\it ibid.}, {\bf 41}, 7326 (1990). 

\bibitem{Marino:1999wg} 
  E.~C.~Marino,
  Quantized skyrmion fields in (2+1)-dimensions,
  Phys.\ Rev.\ B {\bf 61}, 1588 (2000)
  [cond-mat/9910328].

\bibitem{Vlasii:2012kw} 
  N.~D.~Vlasii, C.~P.~Hofmann, F.~J.~Jiang and U.~J.~Wiese,
  Symmetry Analysis of Holes Localized on a Skyrmion in a Doped Antiferromagnet,
  Phys.\ Rev.\ B {\bf 86}, 155113 (2012)
  [arXiv:1205.3677 [cond-mat.str-el]]; 

\bibitem{Vlasii:2014xda} 
  N.~D.~Vlasii, C.~P.~Hofmann, F.-J.~Jiang and U.-J.~Wiese,
  Holes Localized on a Skyrmion in a Doped Antiferromagnet on the Honeycomb Lattice: Symmetry Analysis,
  Annals Phys.\  {\bf 354}, 213 (2014)
  [arXiv:1410.0218 [cond-mat.str-el]].

\bibitem{Acus:2009df} 
  A.~Acus, E.~Norvaisas and Y.~Shnir,
  Baby Skyrmions stabilized by canonical quantization,
  Phys.\ Lett.\ B {\bf 682}, 155 (2009)
  [arXiv:0909.5281 [hep-th]].


\bibitem{Adkins:1983ya} 
  G.~S.~Adkins, C.~R.~Nappi and E.~Witten,
  Static Properties of Nucleons in the Skyrme Model,
  Nucl.\ Phys.\ B {\bf 228}, 552 (1983).

\bibitem{Braaten:1988cc} 
  E.~Braaten and L.~Carson,
  The Deuteron as a Toroidal Skyrmion,
  Phys.\ Rev.\ D {\bf 38}, 3525 (1988).

\bibitem{Irwin:1998bs} 
  P.~Irwin,
  Zero mode quantization of multi - Skyrmions,
  Phys.\ Rev.\ D {\bf 61}, 114024 (2000)
  [hep-th/9804142].

\bibitem{Yabu:1987hm} 
  H.~Yabu and K.~Ando,
  A New Approach to the SU(3) Skyrme Model,
  Nucl.\ Phys.\ B {\bf 301}, 601 (1988).

\bibitem{Hata:2010zy} 
  H.~Hata and T.~Kikuchi,
  Relativistic Collective Coordinate System of Solitons and Spinning Skyrmion,
  Prog.\ Theor.\ Phys.\  {\bf 125}, 59 (2011)
  doi:10.1143/PTP.125.59
  [arXiv:1008.3605 [hep-th]].

\bibitem{Krusch:2005bn} 
  S.~Krusch and J.~M.~Speight,
  Fermionic quantization of Hopf solitons,
  Commun.\ Math.\ Phys.\  {\bf 264}, 391 (2006)
  doi:10.1007/s00220-005-1469-4
  [hep-th/0503067].

\bibitem{Kondo:2006sa} 
  K.~I.~Kondo, A.~Ono, A.~Shibata, T.~Shinohara and T.~Murakami,
  Glueball mass from quantized knot solitons and gauge-invariant gluon mass,
  J.\ Phys.\ A {\bf 39}, 13767 (2006)
  doi:10.1088/0305-4470/39/44/011
  [hep-th/0604006].


\bibitem{Acus:2012st} 
  A.~Acus, A.~Halavanau, E.~Norvaisas and Y.~Shnir,
  Hopfion canonical quantization,
  Phys.\ Lett.\ B {\bf 711}, 212 (2012)
  [arXiv:1204.0504 [hep-th]].



\bibitem{Wilczek:1983cy} 
  F.~Wilczek and A.~Zee,
  Linking Numbers, Spin, and Statistics of Solitons,
  Phys.\ Rev.\ Lett.\  {\bf 51}, 2250 (1983).

\bibitem{Wu:1984kd} 
  Y.~S.~Wu and A.~Zee,
  Comments on the Hopf Lagrangian and Fractional Statistics of Solitons,
  Phys.\ Lett.\ B {\bf 147}, 325 (1984).

\bibitem{Abanov:2000ea} 
  A.~G.~Abanov,
  Hopf term induced by fermions,
  Phys.\ Lett.\ B {\bf 492}, 321 (2000)
  [hep-th/0005150].

\bibitem{Abanov:2001iz} 
  A.~G.~Abanov and P.~B.~Wiegmann,
  On the correspondence between fermionic number and statistics of solitons,
  JHEP {\bf 0110}, 030 (2001)
  [hep-th/0105213].


\bibitem{Bar:2003ip} 
  O.~Bar, M.~Imboden and U.~J.~Wiese,
  Pions versus magnons: from QCD to antiferromagnets and quantum Hall ferromagnets,
  Nucl.\ Phys.\ B {\bf 686}, 347 (2004)
  [cond-mat/0310353].

\bibitem{Jaroszewicz:1985ip} 
  T.~Jaroszewicz,
  Induced Topological Terms, Spin and Statistics in (2+1)-dimensions,
  Phys.\ Lett.\ B {\bf 159}, 299 (1985).





\bibitem{D'Adda:1978uc} 
  A.~D'Adda, M.~Luscher and P.~Di Vecchia,
  A 1/n Expandable Series of Nonlinear Sigma Models with Instantons,
  Nucl.\ Phys.\ B {\bf 146}, 63 (1978).

\bibitem{Piette:1994ug}
  B.~M.~A.~Piette, B.~J.~Schroers and W.~J.~Zakrzewski,
  Multi - Solitons In A Two-Dimensional Skyrme Model,
  Z.\ Phys.\  C {\bf 65}, 165 (1995)
  [arXiv:hep-th/9406160].

\bibitem{Kudryavtsev:1996er}
  A.~E.~Kudryavtsev, B.~Piette and W.~J.~Zakrzewski,
  Mesons, baryons and waves in the baby Skyrmion model,
  Eur.\ Phys.\ J.\  C {\bf 1}, 333 (1998)
  [arXiv:hep-th/9611217].


\bibitem{Varshalovich:1988ye} 
  D.~A.~Varshalovich, A.~N.~Moskalev and V.~K.~Khersonsky,
  {\it Quantum Theory Of Angular Momentum},
  (World Scientific, Singapore, 1988), 528p


\bibitem{Diakonov:1987ty} 
  D.~Diakonov, V.~Y.~Petrov and P.~V.~Pobylitsa,
  A Chiral Theory of Nucleons,
  Nucl.\ Phys.\ B {\bf 306}, 809 (1988).
  doi:10.1016/0550-3213(88)90443-9



\bibitem{amari_un}
Y. Amari, P. Klimas, N. Sawado (to be published). 

\end{thebibliography}
\end{document}